\documentclass[emulateapj,natbib]{emulateapj}

\begin{document}
\title{A First-Look Atmospheric Modeling Study of
the Young Directly-Imaged Planet-Mass Companion, ROXs 42B\lowercase{b}}
\author{
Thayne Currie\altaffilmark{1}, 
Adam Burrows \altaffilmark{2},
Sebastian Daemgen\altaffilmark{1}
}
\altaffiltext{1}{Department of Astronomy and Astrophysics, University of Toronto, 50 St. George St., Toronto, ON, Canada}
\altaffiltext{2}{Department of Astrophysical Sciences, Princeton University}
\begin{abstract}
We present and analyze $J$$K_{s}$$L^\prime$ photometry and our previously published 
$H$-band photometry and $K$-band spectroscopy for ROXs 42Bb, an object 
Currie et al. (2014) first reported as a young directly imaged planet-mass companion.
ROXs 42Bb exhibits IR colors redder than field L dwarfs but consistent with other planet-mass companions.
From the H$_{2}$O-2 spectral index, we estimate a spectral type of L0 $\pm$ 1; weak detections/non-detections of the $CO$ bandheads, 
$Na I$, and $Ca I$ support evidence for a young, low surface gravity object primarily derived from the $H_{2}$(K) index.
ROXs 42Bb's photometry/K-band spectrum are inconsistent with limiting cases of dust-free atmospheres ($COND$) and 
marginally inconsistent with the AMES/DUSTY models and the BT-SETTL models.  However, ROXS 42Bb data 
are simultaneously fit by atmosphere models incorporating several micron-sized dust grains entrained in thick clouds, 
although further modifications are needed to better reproduce the $K$-band spectral shape.
ROXs 42Bb's best-estimated temperature is $T_{eff}$ $\sim$ 1950--2000 $K$, near the low end of the empirically-derived 
range in Currie et al. (2014).  For an age of $\sim$ 1--3 $Myr$ and considering the lifetime of the 
protostar phase, ROXs 42Bb's luminosity of log($L$/$L_{\odot}$) $\sim$ -3.07 $\pm$ 0.07 implies a mass of 9$^{+3}_{-3}$ $M_{J}$,
 making it one of the lightest planetary mass objects yet imaged.
\end{abstract}
\section{Introduction}
ROXs 42Bb is a directly-imaged planetary-mass companion to the 
binary M star and likely $\rho$ Ophiuchus member, ROXs 42B, first reported as such by
 \citet{Currie2014}\footnote{ROXs 42Bb was first identified 
as a point source in the Ophiuchus binary survey from \citet{Ratzka2005}. \citet{Currie2014} 
first publicly identify it as a bound, planet-mass companion.  Independent work announced, accepted for publication, 
 and published after \citet{Currie2014}'s study identify ROXs 42Bb and come to similar conclusions 
about its nature \citep{Kraus2014,Bowler2014}.
}.
The companion 
at a projected separation of $\approx$ 157 $AU$, intermediate between the separations of 
 HR 8799 bcde, GJ 504 b, and HD 95086 b \citep[15--70 $AU$][]{Marois2008,Marois2011,Kuzuhara2013,Rameau2013} 
and much wider-separation planet-mass companions like 1RXJ 1609 B \citep{Lafreniere2008,Lafreniere2010}. 
Based on ROXs 42Bb's near-IR brightness and the shape of its $K$-band spectrum,
\citet{Currie2014} show that ROXs 42Bb's mass is likely below the deuterium-burning limit (M $\sim$ 9$^{+6}_{-3}$ $M_{J}$).

Constraints on ROXs 42Bb's atmospheric properties are limited thus far.
\citet{Currie2014} focused only on simple, empirical comparisons to ROXs 42Bb's 
near-IR photometry and $K$-band spectrum establishing 
it as a low surface gravity, late M/early L dwarf planet-mass companion, reporting a wide range of 
possible temperatures ($T_{eff}$ = 1800--2600 $K$) and surface gravities (log(g) = 4.0 $\pm$ 0.5).
\citet{Bowler2014} add $J$ and $H$ band spectra, independently supporting the conclusion from \citet{Currie2014} 
that ROXs 42Bb is a low surface gravity/mass companion.
But they did not derive a temperature or gravity from template spectra comparisons nor from atmospheric modeling. 
The shape of near-IR spectra and near-IR ($H$/$H$-$K_{s}$) magnitudes/colors 
are sensitive to temperature and/or surface gravity \citep{Luhman2004,Allers2007,Leggett2010,Stephens2009,Currie2013a,Canty2013}.  
However, atmosphere model fits to planet-mass companions focused on the narrow wavelength coverage we presented for 
ROXs 42Bb may yield model parameters that are either unphysical or are inconsistent with data obtained at other 
wavelengths \citep[e.g.][]{Mohanty2007,Barman2011}.  

Data obtained over a wider wavelength baseline, in particular 
including thermal IR data, can better identify the plausible range of substellar object atmosphere parameters \citep{Stephens2009} 
and yield a luminosity estimate.  Comparing this luminosity substellar object evolutionary 
models then yields a mass estimate.
A better determined mass for ROXs 42Bb may also clarify how the companion fits within the context 
of other imaged planetary-mass companions \citep{Currie2014}.  

In this paper, we perform a first-look atmospheric modeling study of ROXs 42Bb.  Our analyses combine
 previously published photometry/spectroscopy with new photometry for ROXs 42Bb and compare these data to 
those for other substellar objects.  We further analyze the ROXs 42Bb spectrum to identify 
 major chemical constituents and place additional constraints on the object's spectral type from 
gravity-independent indices and on its surface gravity from gravity sensitive absorption features.
Using planet/brown dwarf atmosphere models defining limiting cases 
and sophisticated cloud models, we simultaneously
fit photometric and spectroscopic data to better clarify ROXs 42Bb's temperature, surface gravity, 
luminosity and mass.  

\section{Data}

\subsection{New Data}
\subsubsection{Observations and Image Processing}
We reduced $J$, $K_{s}$ and $L^\prime$ data for ROXs 42B obtained from the Keck Observatory Archive (KOA), taken on 
22 June 2011 with the NIRC2 'narrow' camera \citep[9.952 mas pixel$^{-1}$][]{Yelda2010}.
All new data were taken with good adaptive optics corrections 
in classical imaging, and in dither/nod patterns to remove 
the sky background.  The field of view was $\approx$ 10 arc-seconds on a side.
Basic image processing steps 
were identical to those carried out for previously published
ROXs 42Bb photometry \citep{Currie2014}.  
Briefly, we first sky-subtracted each image from a 
median combination of images obtained at other dither positions.
Then, we identified and removed hot/cold/bad pixels, corrected each image for distortion using 
the \textit{nirc2dewarp.pro} \textit{IDL} routine, copied 
 each to larger blank image, and registered the images to a common center \citep{Currie2012a,Currie2012b}.
We then subtracted off a 2D radial profile of the primary to remove the halo light from each 
image.  After these steps, we median-combined together our set of images and rotated the 
combined image to the north-up position ($PA_{north}$ = -0.252$^{o}$; Yelda et al. 2010).

\begin{figure}
\centering
\includegraphics[trim=25mm 0mm 25mm 0mm, clip,scale=0.3]{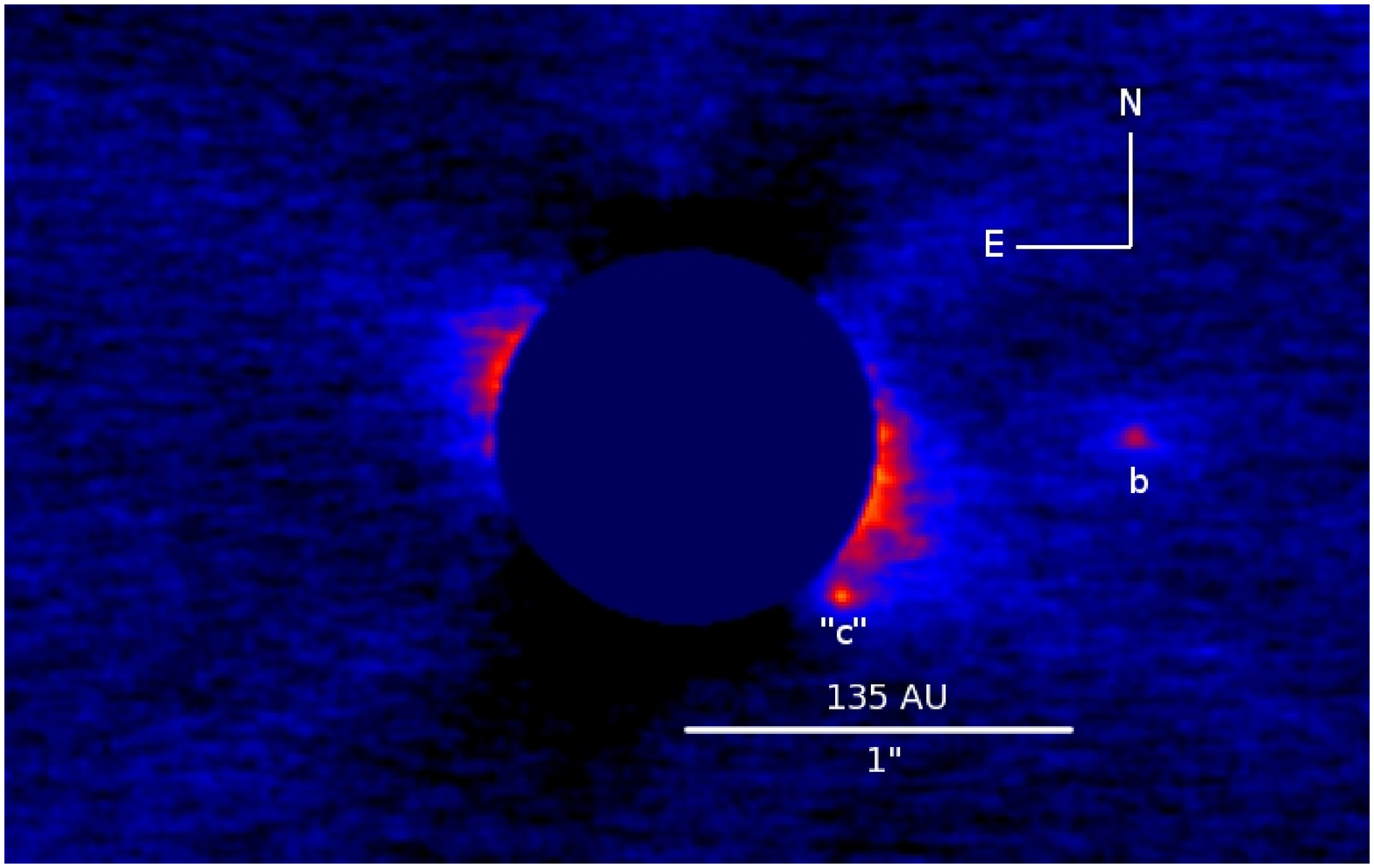}
\includegraphics[trim=25mm 0mm 25mm 0mm, clip,scale=0.3]{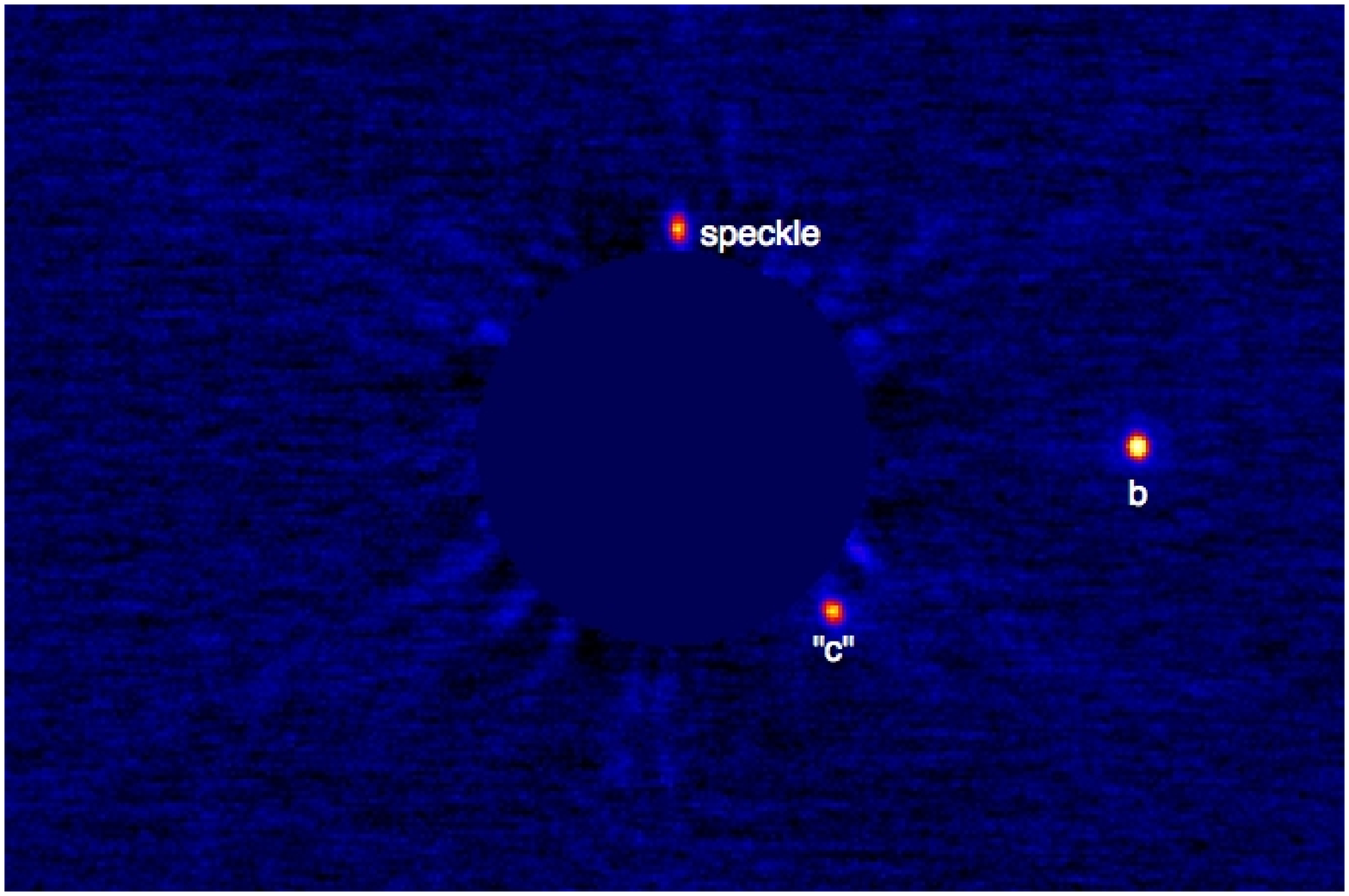}
\includegraphics[trim=25mm 0mm 25mm 0mm, clip,scale=0.3]{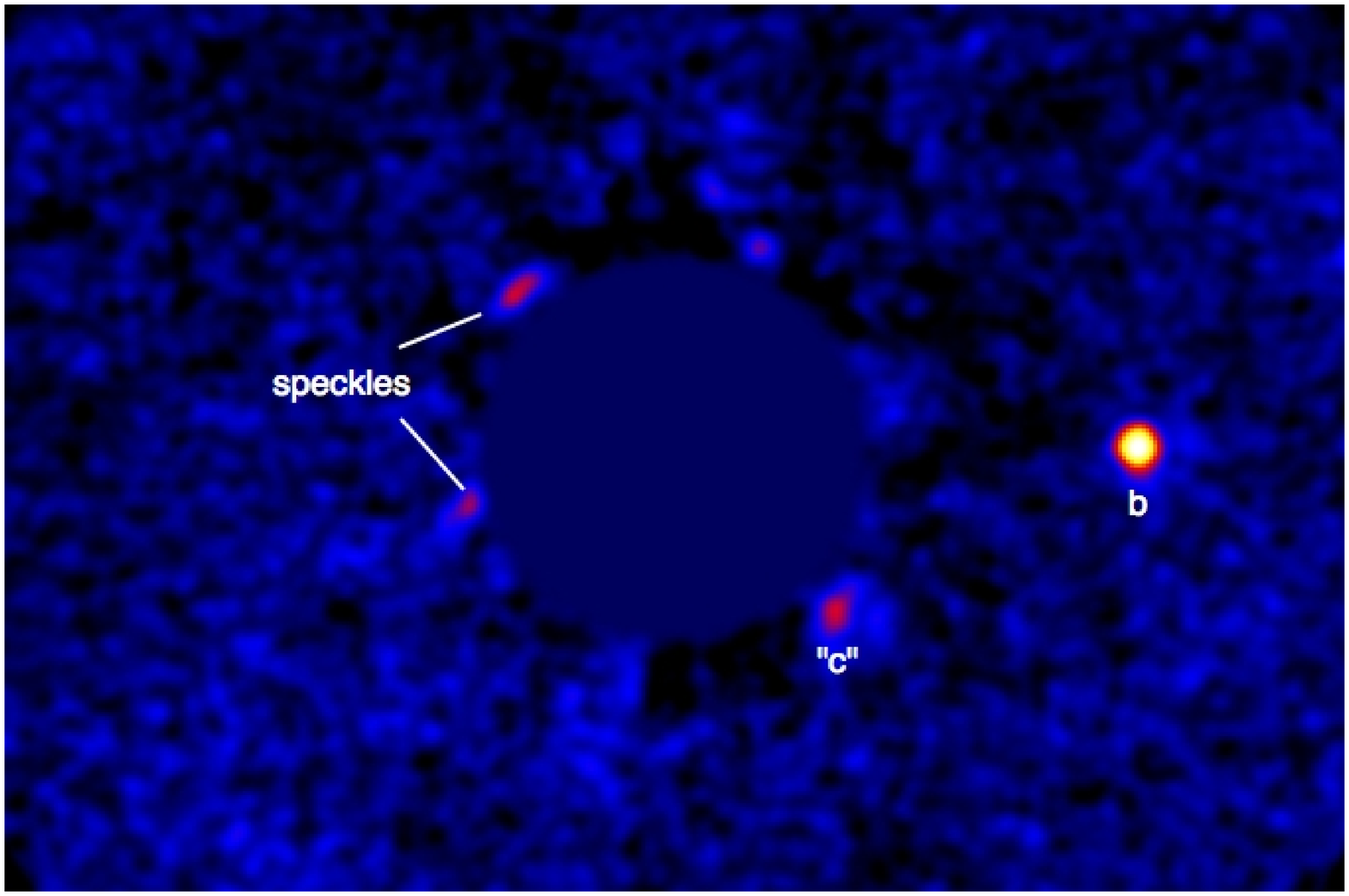}
\caption{Images of ROXs 42B from $J$ (left), $K_{s}$ (middle), and 
$L^\prime$ (right) 2011 NIRC2 data.  ROXs 42Bb and the background object 
``ROXs 42Bc" are detected in each image. }
\label{images}
\end{figure}

Figure \ref{images} displays the combined $J$ (left), $K_{s}$ (middle), and $L^\prime$ images, revealing ROXs 42Bb 
 at an angular separation of $\approx$ 1\farcs{}17 and position angle 
of $\approx$ 270$^{o}$.  Additionally, each data set reveals the second candidate companion, ``ROXs 42B c", 
at roughly 0\farcs{}5 separation, although the $L^\prime$ detection is noticeably weaker than detections 
at $J$ and $K_{s}$.  ROXs 42B is defined as a close binary ($\Delta$$\theta$ $\approx$ 0\farcs{}05) identified from lunar occultation 
data \citep[e.g.][]{Simon1995}; however, we fail to resolve the two components \citep[see also][]{Currie2014,Kraus2014}.

\subsubsection{Photometry}
We largely follow previous methods used to extract photometry for the ROXs 42B system 
 as described in \citet{Currie2014}.  For each data set, we measure the integrated signal for 
 both the primary (in a median-combined registered image) and the companions 
(in a median-combined radial-profile subtracted image) in an aperture sized to the 
 full-width half-maximum of the primary star and define and subtract off the background signal in 
a surrounding annulus.  As a separate check, we extracted photometry assuming 
a zero background signal and with slightly different sized apertures.
  Our photometric uncertainties incorporate 
the intrinsic SNR of the companions, the SNR of the primary star, and differences in brightness 
measurements adopting different aperture sizes and background treatments.

These procedures yield a companion-to-primary contrast of 
$\Delta$J = 7.00 $\pm$ 0.11, $\Delta$$K_{s}$ = 6.33 $\pm$ 0.06, and $\Delta$$L^\prime$ = 5.64 $\pm$ 0.06
for ROXs 42Bb.  For the putative ``ROXs 42Bc" object, we derive $\Delta$J=6.69 $\pm$ 0.12, 
$\Delta$$K_{s}$ = 6.78 $\pm$ 0.07, and $\Delta$$L^\prime$ = 6.76 $\pm$ 0.20.  
Our K-band photometry
for ROXs 42Bb agrees with that derived from older Keck/NIRC2 data reported in \citet{Currie2014}.
\citet{Kraus2014}'s $J$ and $L^\prime$ photometry agrees with our values for ROXs 42Bb.  One of 
their $K_{s}$ band estimates agree with ours while the other two measurements of ROXs 42Bb are significantly too faint 
or bright and in fact inconsistent with one another given their reported photometric errors and adopted limits 
on stellar variability, discrepancies potentially due to their aperture photometry methods (see Section 2.2).  

To flux-calibrate the companions' photometry in $J$ and $K_{s}$, we simply adopt the 2MASS measurements as reported 
before: 9.906 $\pm$ 0.02 and 8.671 $\pm$ 0.02.  To estimate the $L^\prime$ brightness of ROXs 42B, we adopt the 
intrinsic colors for young pre-main sequence stars listed in \citet{Pecaut2013}, assume that the $K_{s}$-$L^\prime$ 
color is between the predicted $K_{s}$-$W1$ and $K_{s}$-$W2$ 2MASS/WISE colors ($\lambda_{o}$ ($W1$,$W2$) = 3.4, 4.6 $\mu m$) or 
$K_{s}$ -$L^\prime$ = 0.14, and assume a 0.05 mag uncertainty in $K_{s}$-$L^\prime$ color from the difference between the 
two 2MASS/WISE colors\footnote{While ROXs 42B is a binary with unequal mass components, the brightness ratio at 
$K_{s}$ band is $\sim$ 2 to 1, and the $K_{s}$-W1 and $K_{s}$-W2 colors for young M0--M5 dwarfs varies by less 
than 0.1 and 0.25 magnitudes.  
So the $K_{s}$-$L^\prime$ color should reflect that of the brighter component.  There is no evidence of 
ROXs42B being variable in brightness.}
Adopting the \citet{Cardelli1989} reddening relations and assuming $A_{V}$ = 1.9 \citep{Currie2014} 
then yields a predicted $L^\prime$ magnitude of 8.42 $\pm$ 0.05.  This value also agrees with a 
straight or weighted average of ROXs 42B's measured K$_{s}$-W1 and K$_{s}$-W2 colors (8.40--8.43 $\pm$ 0.03).
To then estimate the dereddened absolute magnitudes for ROXs 42Bb we assume a distance of 135 $\pm$ 8 $pc$ 
\citep{Mamajek2008} and the same values for the extinction and reddening law.

\subsection{Previously Published Data}
To these data we add VLT/SINFONI spectra and $H$ band photometry previously published in \citet{Currie2014}.  
We measure the primary-to-companion contrast for ROXs 42Bb as 
$\Delta$$H$ = 6.86 $\pm$ 0.05.  Given the 2MASS $H$-band photometric measurement of m$_{H}$ = 9.017 $\pm$ 0.020, 
we derive an apparent magnitude of 15.88 $\pm$ 0.06 and
a dereddened absolute magnitude of 9.87 $\pm$ 0.14.

As noted in \citet{Currie2014} the putative ``ROXs 42Bc" is of comparable brightness in $H$ band.  Quantitatively, 
we derive a contrast of $\Delta$H = 6.94 $\pm$ 0.09 and an apparent magnitude of 
15.96 $\pm$ 0.09.  These values significantly disagree with the contrast estimate by \citet{Kraus2014} who 
claim $\Delta$H = 6.20.  
With the background signal largely removed from a radial profile subtraction or explicitly measured 
in a median-combined frame, ``ROXs 42Bc" clearly cannot be $\sim$ twice as bright as ROXs 42Bb  
\footnote{The implied relative brightness of ``ROXs 42Bc" \textit{is} higher
in raw counts from median-combined, registered images without proper background subtraction, with peak 
values of roughly $\sim$ 1225 cts. vs. 810 cts for ROXs 42Bb.  However, the background signal around ``ROXs 42Bc" is 
500 cts higher.}.  

Table \ref{newphot} summarizes our photometric measurements for both ROXs 42Bb and ``ROXs 42Bc".
While both companions have similar $H$-band contrasts and apparent magnitudes, their observed colors 
are quite different.  In particular, ROXs 42Bb has a $J$-$L^\prime$ color of 2.86, significantly redder 
than that of ``ROXs 42Bc" (1.42), a difference largely owing to the latter's significantly fainter 
$L^\prime$ flux density.  

Additional arguments show that ``ROXs 42Bc" is likely a reddened background object.  Besides being 
significantly fainter at $L^\prime$, it is about 0.45 mags fainter at $K_{s}$ band and $\sim$ 0.31 mags brighter at $J$.  
If the line-of-sight extinction between us and ``ROXs 42Bc" is in the $A_{V}$ = 4-6 range consistent
with it being behind the filamentary structure containing ROXs 42B, then the object dereddens to the colors 
expected for an early-type star.  For example, assuming $A_{V}$ = 4.5 yields m($J$)= 15.33 $\pm$ 0.12, 
m($H$) = 15.11 $\pm$ 0.09, m($K_{s}$) = 14.94 $\pm$ 0.07 and m($L^\prime$) = 14.93 $\pm$ 0.20, consistent with the 
colors for an A0 to mid F star \citep[see][]{Pecaut2013,Currie2010b}.  
Thus, as previously suggested in \citet{Currie2014} and argued independently by \citet{Kraus2014}, our analysis 
shows that ``ROXs 42Bc" is likely not a second planetary-mass companion orbiting ROXs 42B.

\section{Direct Constraints on ROXs 42B\lowercase{b}'s Atmosphere}
\subsection{IR Colors}
To compare the near-to-mid IR photometric properties of ROXs 42Bb with
those for other substellar objects, we follow similar analysis in
\citet{Currie2013a}, using the \citet{Leggett2010} sample of field
brown dwarfs with spectral classes between M7 and T5.  To these objects,
we add the directly-imaged planetary-mass companions/candidates
listed in their Table 4 using photometric measurements listed in
their Table 5.  For clarity, we do not include young objects with inferred 
masses above the deuterium-burning limit: these tend to be much more luminous 
in each infrared filter than ROXs 42Bb.

\begin{figure}
\centering
\includegraphics[scale=0.4]{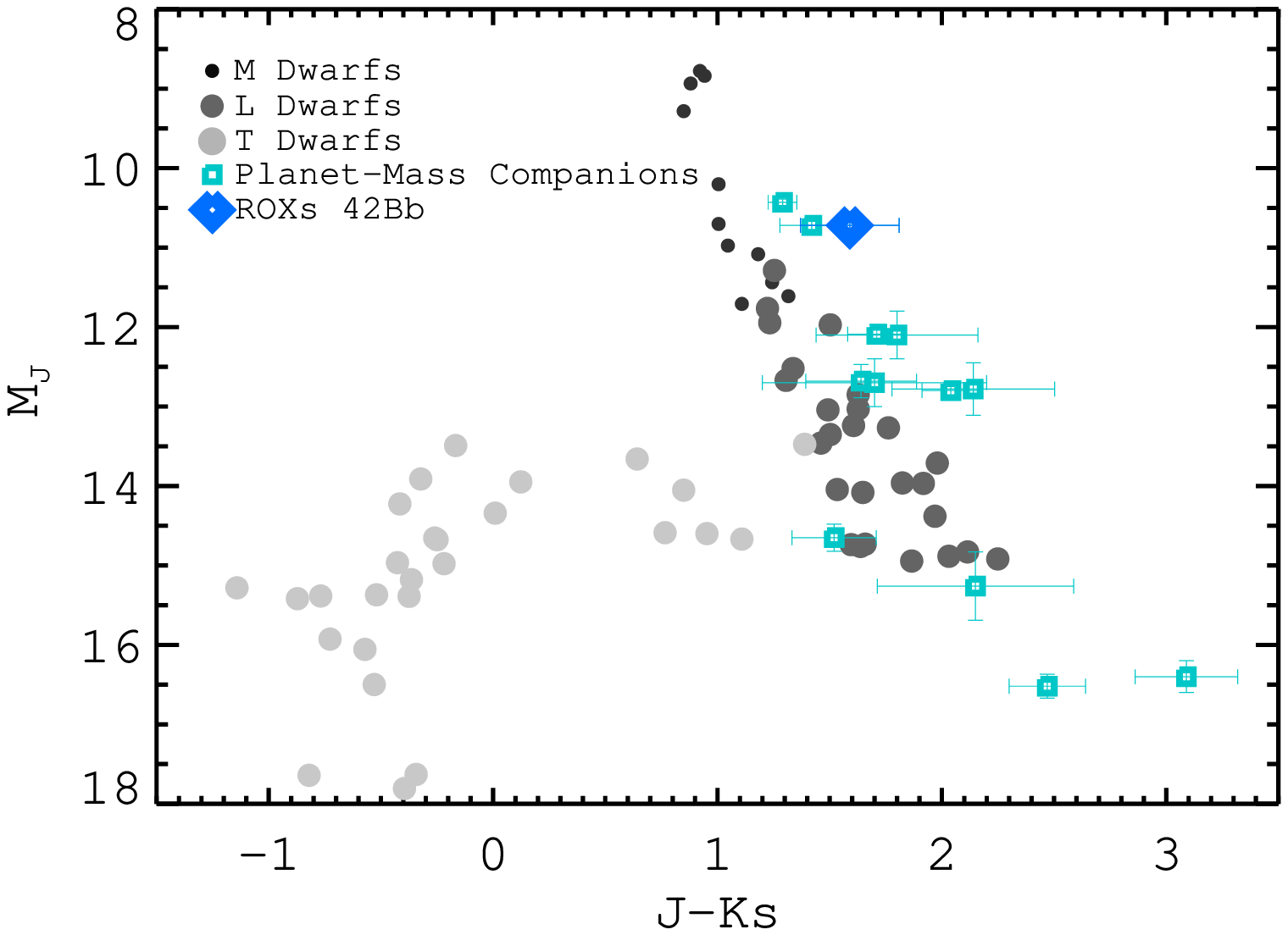}
\includegraphics[scale=0.4]{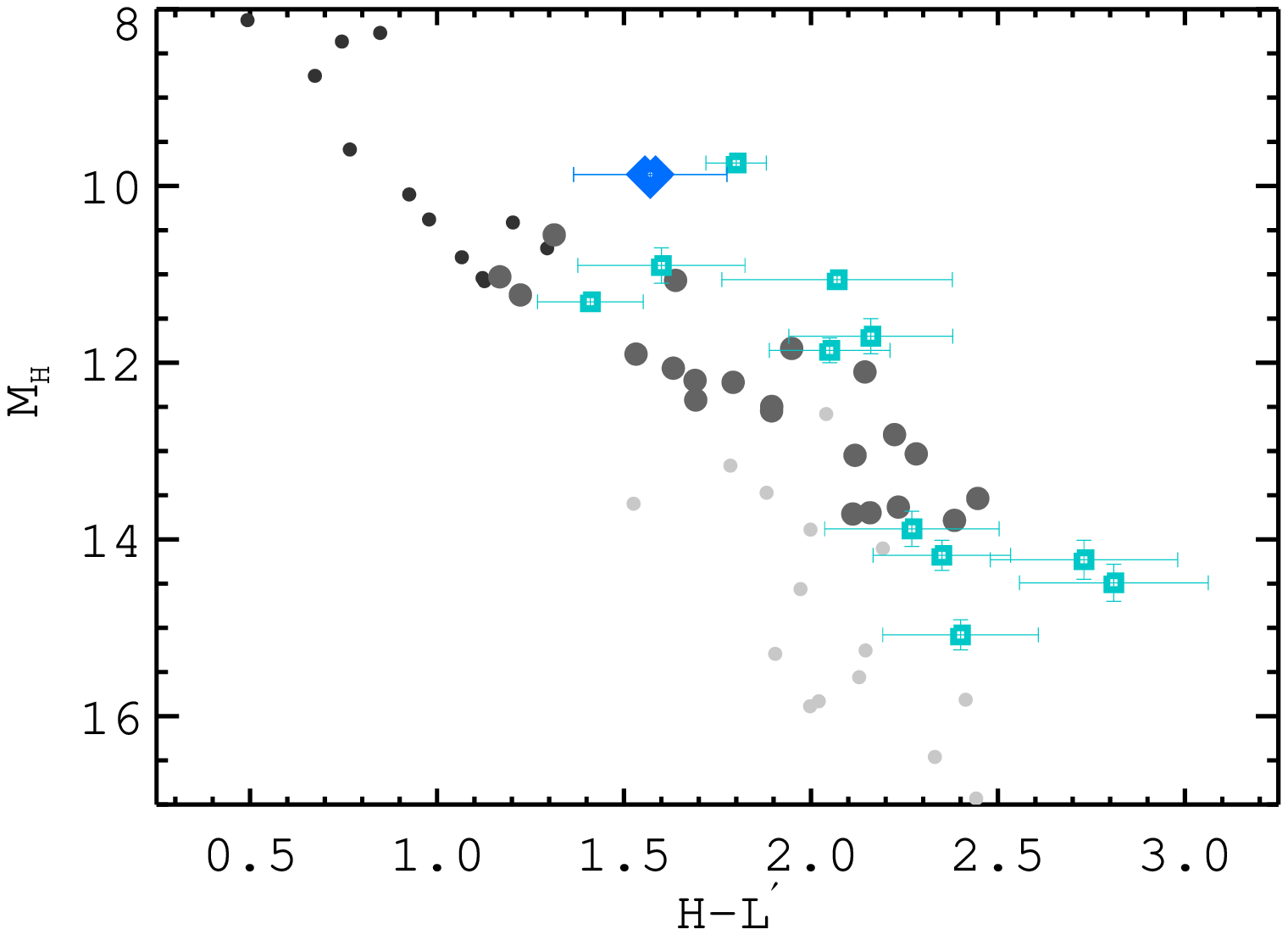}
\caption{(Left) $J$/$J$-$K_{s}$ and $H$/$H$-$L^\prime$ color-magnitude diagrams comparing 
ROXs 42Bb's photometry (blue diamonds) to MLT dwarfs from \citet{Leggett2010} and other 
planetary-mass companions (aqua squares).}
\label{colcol}
\end{figure}

Figure \ref{colcol} compares the $J$/$J$-$K_{s}$ and $H$/$H$-$L^\prime$ color-magnitude
position of ROXs 42Bb (blue diamond) to those positions for
older, field brown dwarfs (black/grey dots) and other planetary-mass companions (aqua squares).
\citet{Currie2014} found that the near-IR $H$/$H$-$K_{s}$ color-magnitude diagram position 
for ROXs 42Bb, like that for directly-imaged planets $\beta$ Pic b and HR 8799 bcde \citep{Marois2008,Currie2011a,Currie2013a},
 was discrepant compared to the field MLT dwarf sequence.  The same trends 
persist when we consider $J$ and $L^\prime$ data, as ROXs 42Bb is $\sim$ 0.5 mags redder than the 
field sequence in both $J$-$K_{s}$ and $H$/$H$-$L^\prime$.   

Compared to the planet-mass companions plotted, ROXs 42Bb's $J$/$J$-$K_{s}$, $H$/$H$-$L^\prime$, and 
$H$/$H$-$K_{s}$ \citep{Currie2014} color-magnitude diagram position 
bears the strongest resemblance to 
6--16 $M_{J}$ M/L transition objects 
GSC 06214B and USco CTIO 108 B \citep{Ireland2011,Bejar2008}.  These objects have masses of 
$\sim$ 6--16 $M_{J}$, and spectral types of M9.5--L0. Given their slightly older but qualitatively 
similar ages of $\approx$ 5--10 $Myr$ \citep{Pecaut2012,Bowler2011}, we expect they should serve as 
good indicators for ROXs 42Bb's spectral type and inferred mass.

\subsection{Spectral Analysis}
\begin{figure}
\centering
\includegraphics[scale=0.6]{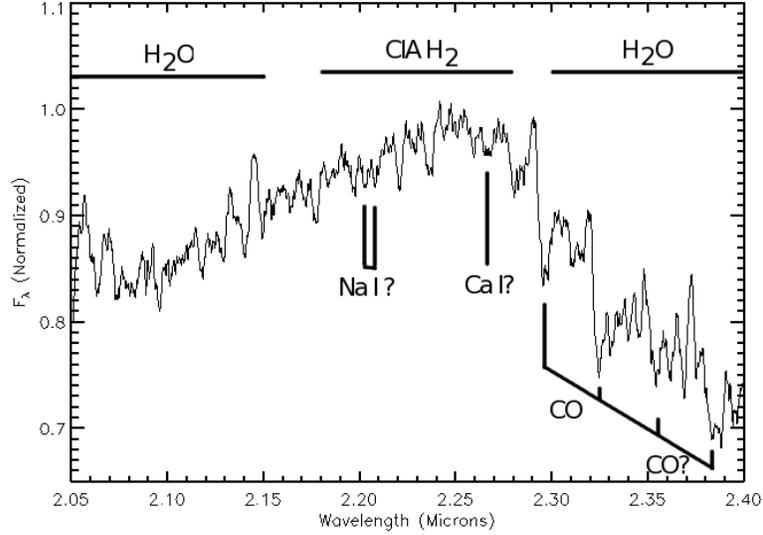}
\caption{Smoothed, binned ROXs 42Bb continuum with detected broad molecular features -- $H_{2}$O and 
the gravity sensitive collisionally-induced $H_{2}$ feature -- and the CO bandhead along 
with the locations of Na I and Ca I lines that serve as secondary gravity indicators where 
we do not have strong detections.}
\label{molid}
\end{figure}

In \citet{Currie2014}, we focused on simple, empirical comparisons to the ROXs 42Bb 
spectral shape to estimate its spectral type and assess evidence for low surface 
gravity \citep[see also][]{Bowler2014}.  Here, we further analyze the spectrum (Figure \ref{molid}) to 
estimate its spectral type from gravity independent (or weakly dependent) indices \citep{Allers2013,Bonnefoy2013}, 
identify chemical species present,
and corroborate evidence for low surface gravity from alternate diagnostics \citep{Canty2013}.

\subsubsection{Spectral Type}
To estimate the spectral type of ROXs 42Bb, we compute the H$_{2}$0-2 index, defined by 
\citet{Slesnick2004} as the 2.04 $\mu m$ to 2.15 $\mu m$ flux ratio.  This index appears 
well correlated with spectral type over the range enclosing ROXs 42Bb's likely value and for 
substellar objects in the Orion Nebula Cluster, which are roughly coeval with ROXs 42Bb.
As this is an estimate of the strength of the broad $H_{2}$O absorption feature at $\approx$ 2 $\mu m$, not 
a narrow line feature, we compute the index for a smoothed, binned version of the spectrum that we later use for 
atmospheric modeling ($R$ $\sim$ 650, see Sect. 4)\footnote{We derive similar estimates for the raw spectrum.}.

We measure ROXs 42Bb's H$_{2}$O-2 index to be [H$_{2}$0] $\sim$ 0.87.
To estimate a spectral type from this index, we consider identical measurements for young 
substellar objects in Upper Scorpius \citep{Lodieu2008}, the relationship 
mapping between these two quantities in \citet{Bonnefoy2013}, and the polynomial fit from \citet{Allers2013}.
We then take the average of these values as our spectral type and the dispersion in these values 
as our uncertainty.

The \citet{Allers2013} polynomial fit yields a spectral type of L0, whereas the \citet{Bonnefoy2013} 
relationship yields M9 and \citet{Lodieu2008} empirical scale yields L0--L1.
From these comparisons, we then estimate ROXs 42Bb's spectral type to be L0 $\pm$ 1 subclass.  
Thus, our estimate for ROXs 42Bb's spectral type is consistent with simple empirically-based estimates 
from \citet{Currie2014} and \citet{Bowler2014} of M8--L0 and L1 $\pm$ 1, respectively.

\subsubsection{Chemical Composition and Gravity-Sensitive Features}
The spectra of late M/early L dwarfs in $K$-band are characterized by 
strong H$_{2}$O absorption from the blue end of the bandpass through $\approx$ 2.1 $\mu m$, 
collisionally-induced absorption of $H_{2}$ (CIA $H_{2}$) at 2.18--2.28 $\mu m$ \citep{Kirkpatrick2006,Saumon2012}, 
prominent CO lines starting at 2.29 $\mu m$, and H$_{2}$O absorption longwards of 2.3 $\mu m$.
Field objects near the M/L transition have weaker Na I doublet lines at 2.206 and 2.209 $\mu m$ and Ca I triplet emission 
(2.261, 2.263, and 2.265 $\mu m$) than earlier spectral types \citep[earlier than M7;][]{Cushing2005}. 
These features may weaken further for younger, lower surface gravity counterparts \citep{Canty2013}.

In the ROXs 42Bb spectrum (Figure \ref{molid}), we see clear evidence for H$_{2}$O absorption at the blue and red ends, 
similar to that seen in both the $H$ and $K$-band spectrum presented in \citet{Bowler2014}. 
Our spectrum also exhibits weak CIA $H_{2}$ responsible for the rather red 2.17--2.24 $\mu m$ slope indicative of 
very low surface gravity planet-mass companions \citep[c.f.][]{Kirkpatrick2006,Canty2013,Currie2014}.
Likewise, ROXs 42Bb exhibits CO bandhead structure at $\lambda$ $>$ 2.29 $\mu m$.  The spectrum 
is particularly noisy near the bandpass limits and thus we do not look for evidence of Ca I 
absorption at $\sim$ 1.98 $\mu m$.

To augment our analysis of CIA $H_{2}$ in \citet{Currie2014}, we look for evidence of other, 
more weakly gravity sensitive features Na I/2.21 $\mu m$ and Ca I/2.26 $\mu m$
studied in \citet{Canty2013}.  While the CO bandhead is clearly detected, the features are
particularly weak.  Using the \textit{IRAF} task \textit{splot}, we calculate an equivalent 
width of $W_\mathrm{CO(2-0)} \sim 8\pm2$\AA\ at 2.29 $\mu m$, comparable to or smaller than any 
young object studied in \citet{Canty2013} and much smaller than equilvalent widths for 
field late M/early L dwarfs.  We do see local minima at the positions of the Na I and 
Ca I features with equivalent widths of $\approx$ 2 \AA\ and 1 \AA.  However, 
they are difficult to distinguish from the noise in the spectrum and 
thus their detections are marginal at best.  
These estimates consistent with results for ONC M/L transition objects 
but inconsistent/marginally consistent with measurements for field 
objects \citep{Canty2013}.  Thus, the line strengths of these secondary gravity diagnostics 
agree with our general conclusion based on the H$_{2}$(K) index: ROXs 42Bb is a low 
surface gravity substellar object.

\section{Atmospheric Modeling}

\textbf{Atmosphere Models Considered} -- To explore ROXs 42Bb's atmospheric properties, we 
compare the object's broadband photometry 
and $K$-band spectrum to predictions from planet atmosphere models adopting a range of 
effective temperatures, surface gravities, and cloud/dust prescriptions.
Table \ref{atmosparam} summarizes our parameter space.
We consider two limiting cases for dust/clouds 
in model atmospheres, the cloud/dust free COND models \citep{Baraffe2003} and the 
AMES/DUSTY models which assume that submicron-sized dust grains are suspended everywhere 
in the atmosphere.  Then we compare the ROXs 42Bb data to the BT-Settl model atmospheres \citep{Allard2012}
and our own \citep{Currie2013a}, both of which adopt cloud models.
The COND and DUSTY models cover a wide range in temperature and gravity ($T_{eff}$ = 1600--2800 $K$ where 
$\Delta$$T_{eff}$ = 100 $K$, log(g) = 3.5--5 where $\Delta$(log(g))=0.5).  The BT-Settl model grid 
 is smaller, as is our own model grid (hereafter referred to as ``the Burrows grid").

\textbf{Fitting Method} -- To model photometry, we simply compare the measured flux densities with predicted ones 
from the model spectra convolved with each filter function.  The COND, DUSTY, and BT-Settl grid resolutions are at significantly 
higher resolution than the data ($R$ $\approx$ 4000), whereas the Burrows grid is at lower resolution 
($R$ $\sim$ 650).  To provide more direct quantitative comparisons amongst the model fits, we smooth 
and rebin the other atmosphere models to the Burrows model grid.  Similarly, we rebin the extracted ROXs 42Bb spectrum
to the Burrows grid resolution\footnote{We found nearly identical modeling results for the COND, DUSTY, and BT-Settl 
grid if we simply degraded their resolution to that of the data.}.    Our modeled wavelength range is 1.95 to 2.45 $\mu m$.

To conservatively estimate errors in each spectral channel, we model and subtract 
off the pseudo-continuum, compute the rms noise in a moving box centered on each channel.  
Furthermore, we consider systematic errors in the spectral extraction, defined as 
differences in the spectrum's flux density as a result of various measurements of the spatially variable background.
Finally, we 
consider errors in the spectra measurements from uncertainties in dereddening the spectrum that are ultimately 
due to uncertainties in the spectral type of ROXs 42B \citep[K5--M1; see footnote in][]{Currie2014}.
The signal-to-noise ratio of the binned spectrum varies from a peak of $\approx$ 25 to $\sim$ 5 for the bluest and 
reddest channels.  Thus, our model fitting is by far most sensitive to matches near the peak of the spectrum 
(2.1--2.2 $\mu m$) than the edges, which could be affected by telluric contamination.

We treat the planet radius as a free parameter, 
varying it between 0.9 $R_{J}$ and 3 $R_{J}$.  We identify the 
 set of models formally consistent with the spectra or photometry at the 95\% confidence limits given 
the number of degrees of freedom \citep{Currie2011a}. 
Although we adopt the dereddened absolute magnitude to compare photometry with each model, we adopt only the 
\textit{measured} photometric error (listed in fourth column of Table \ref{newphot}) for model fitting and incorporate 
our distance uncertainty in the uncertainty estimate for the planet radius.
We derive results for photometry and $K$-band spectroscopy separately and then compare them to identify 
the subset of models consistent with both.

\subsection{Photometry Fitting Results}
\begin{figure}
\centering
\includegraphics[scale=0.5]{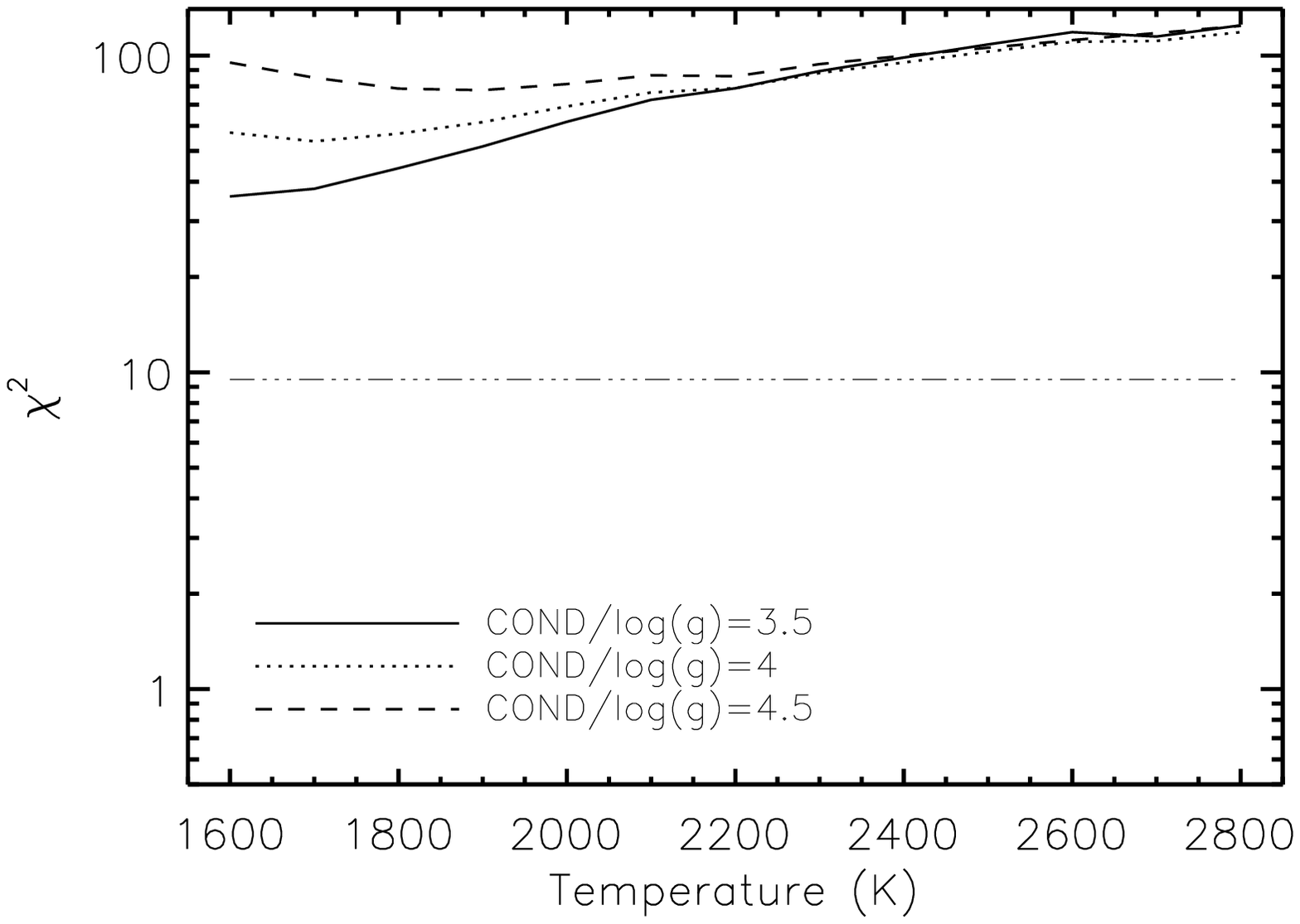}
\includegraphics[scale=0.5]{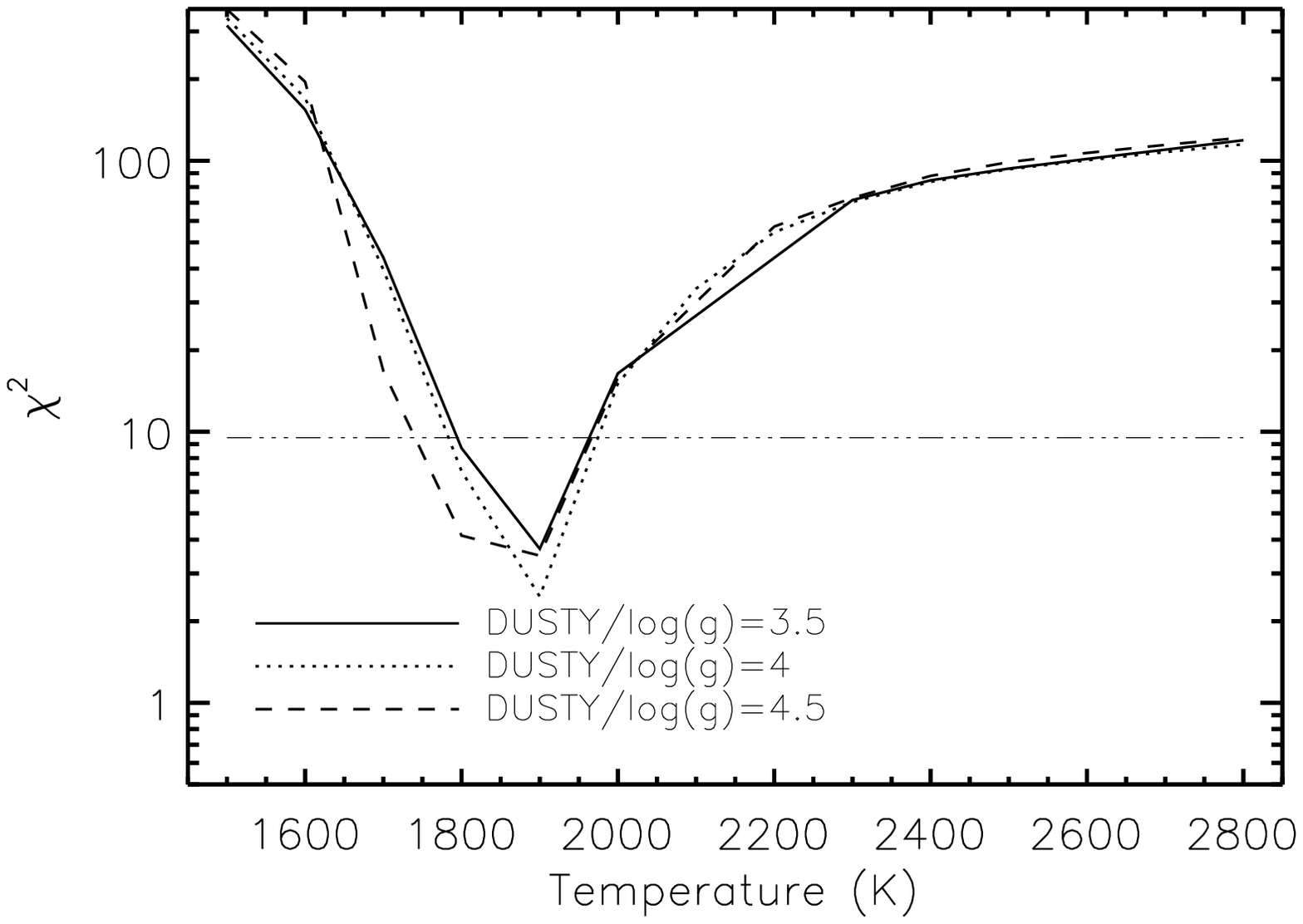}
\includegraphics[scale=0.5]{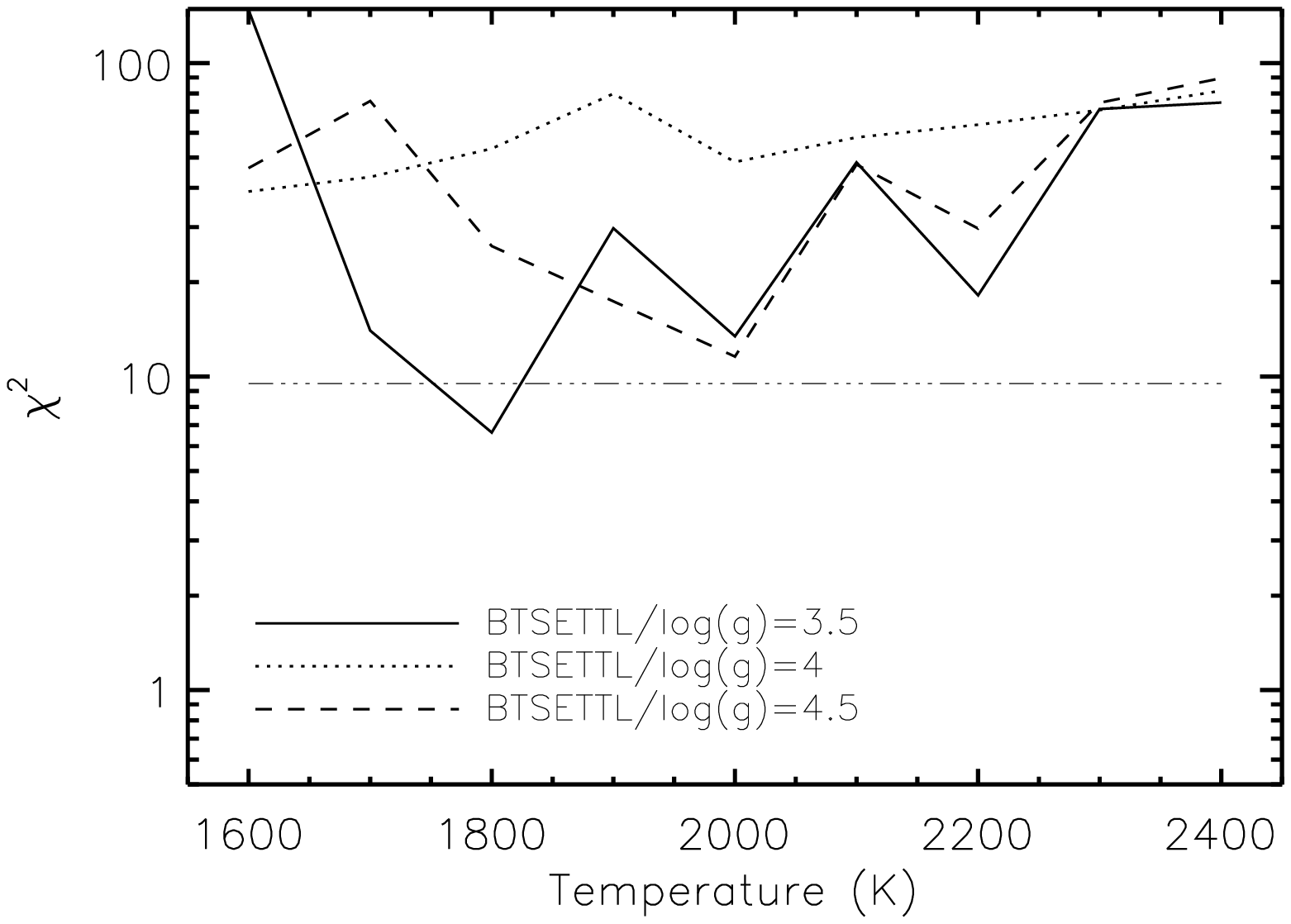}
\includegraphics[scale=0.5]{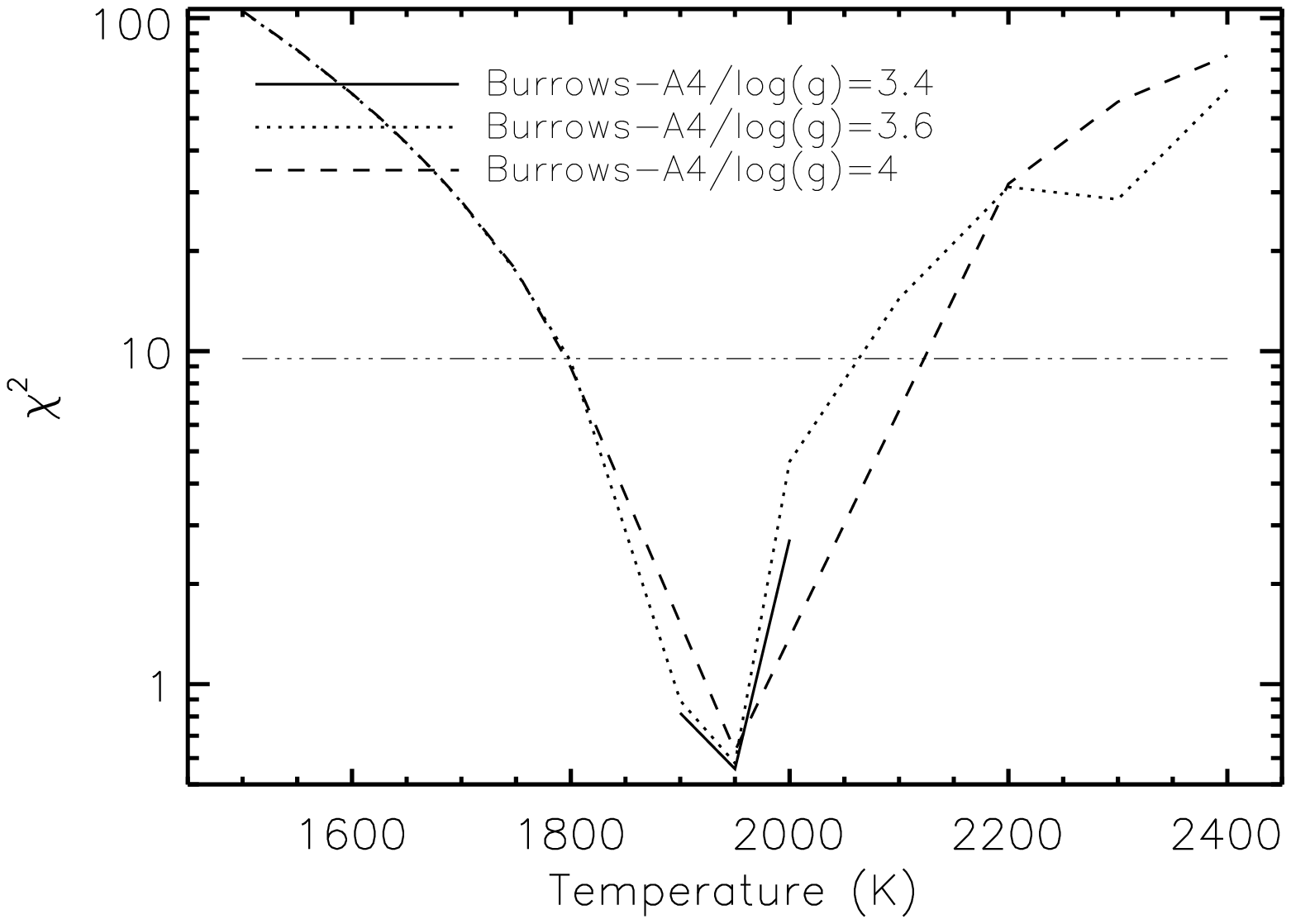}
\caption{The $\chi^{2}$ distributions for fitting ROXs 42Bb photometry with the $COND$ (top-left), 
$DUSTY$ (top-right), $BT-Settl$ (bottom-left), and Burrows (bottom-right) atmosphere models.  The horizontal 
dash-three dotted line identifes the $\chi^{2}$ limit below which models agree with the data to within 
the 95\% confidence limit.
For clarity, we 
do not display all models of different surface gravity for the Burrows grid as the general trends match those for 
models with different surface gravities.
}
\label{photfit}
\end{figure}

\begin{figure}
\centering
\includegraphics[scale=0.5]{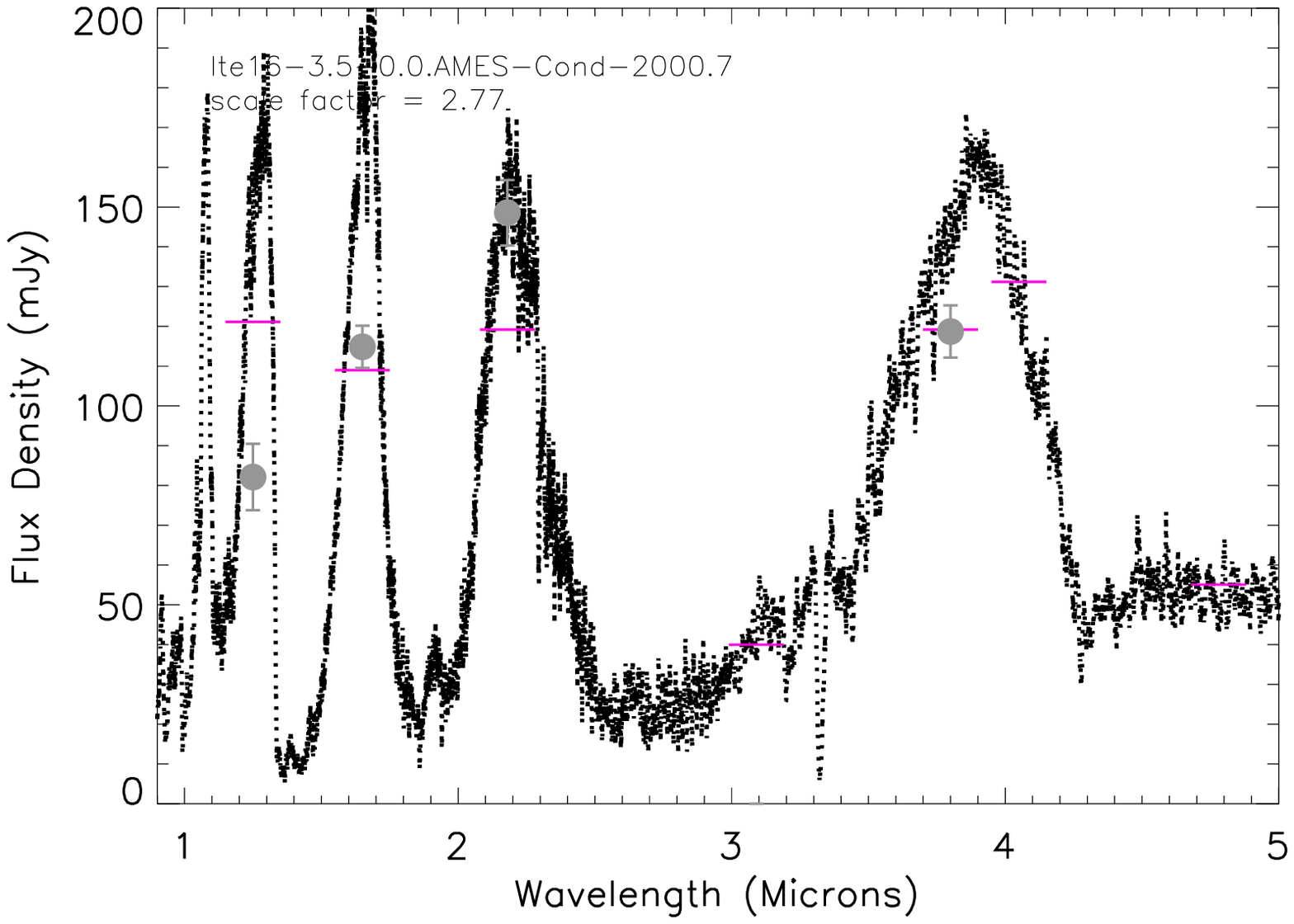}
\includegraphics[scale=0.5]{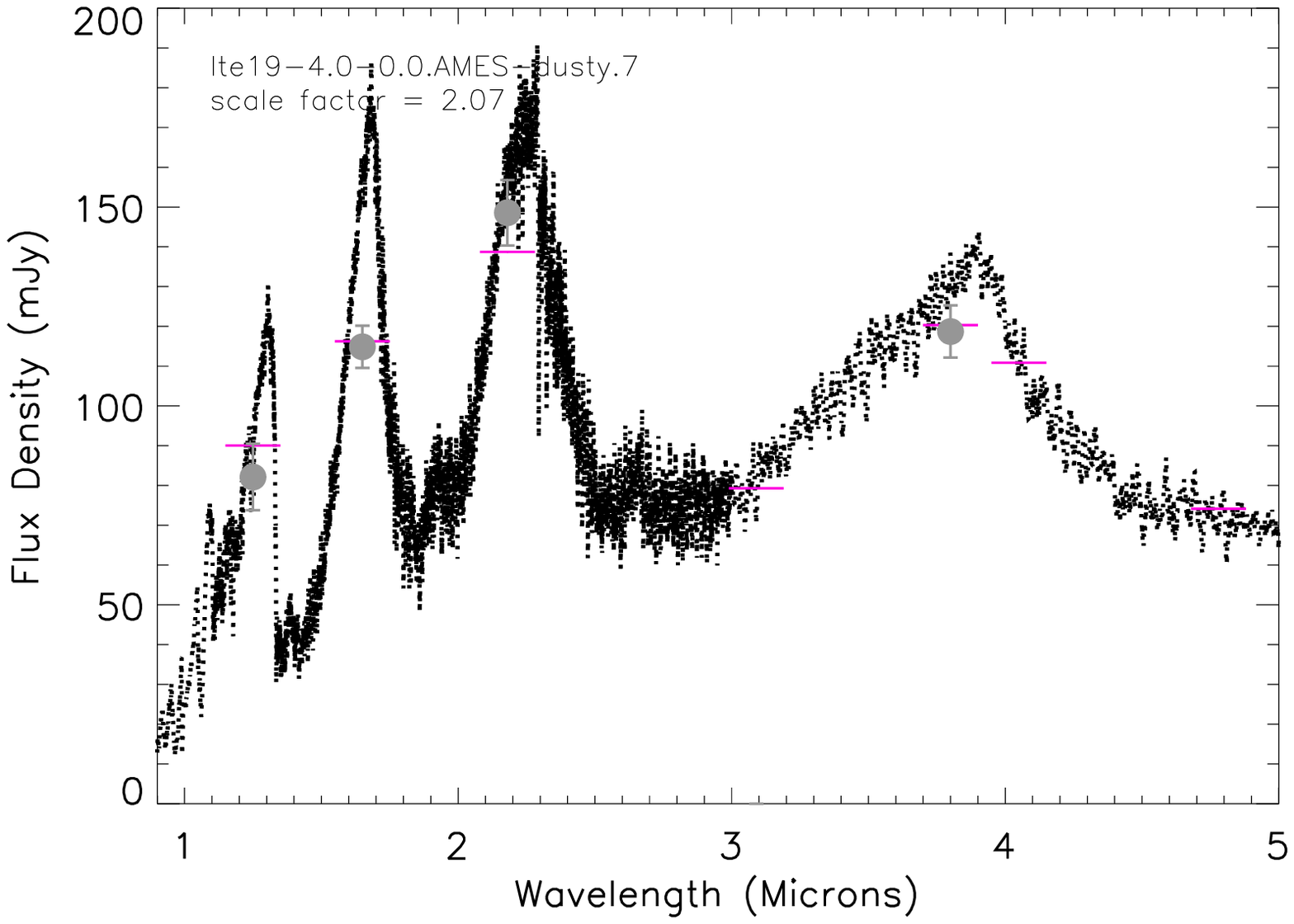}
\includegraphics[scale=0.5]{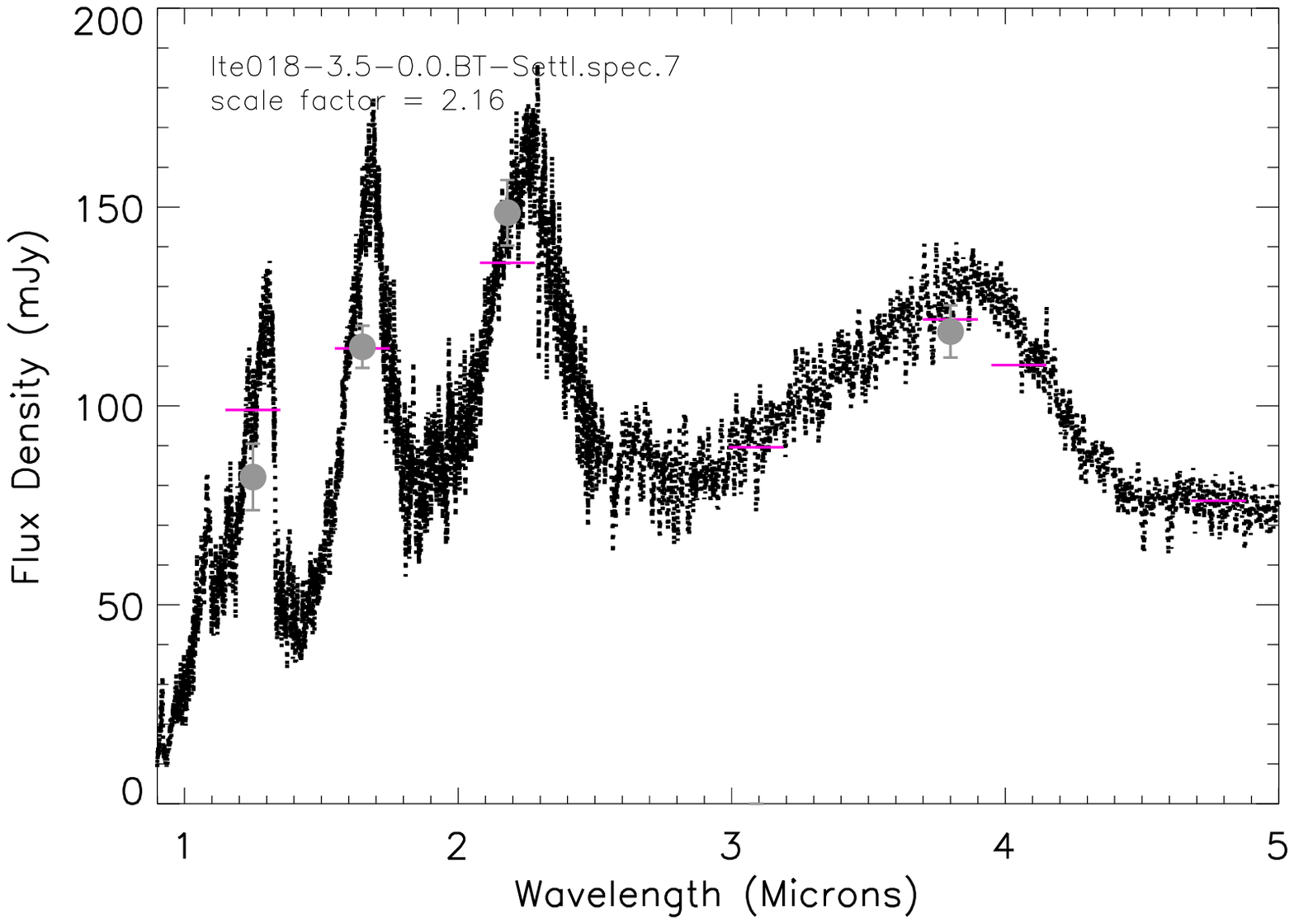}
\includegraphics[scale=0.5]{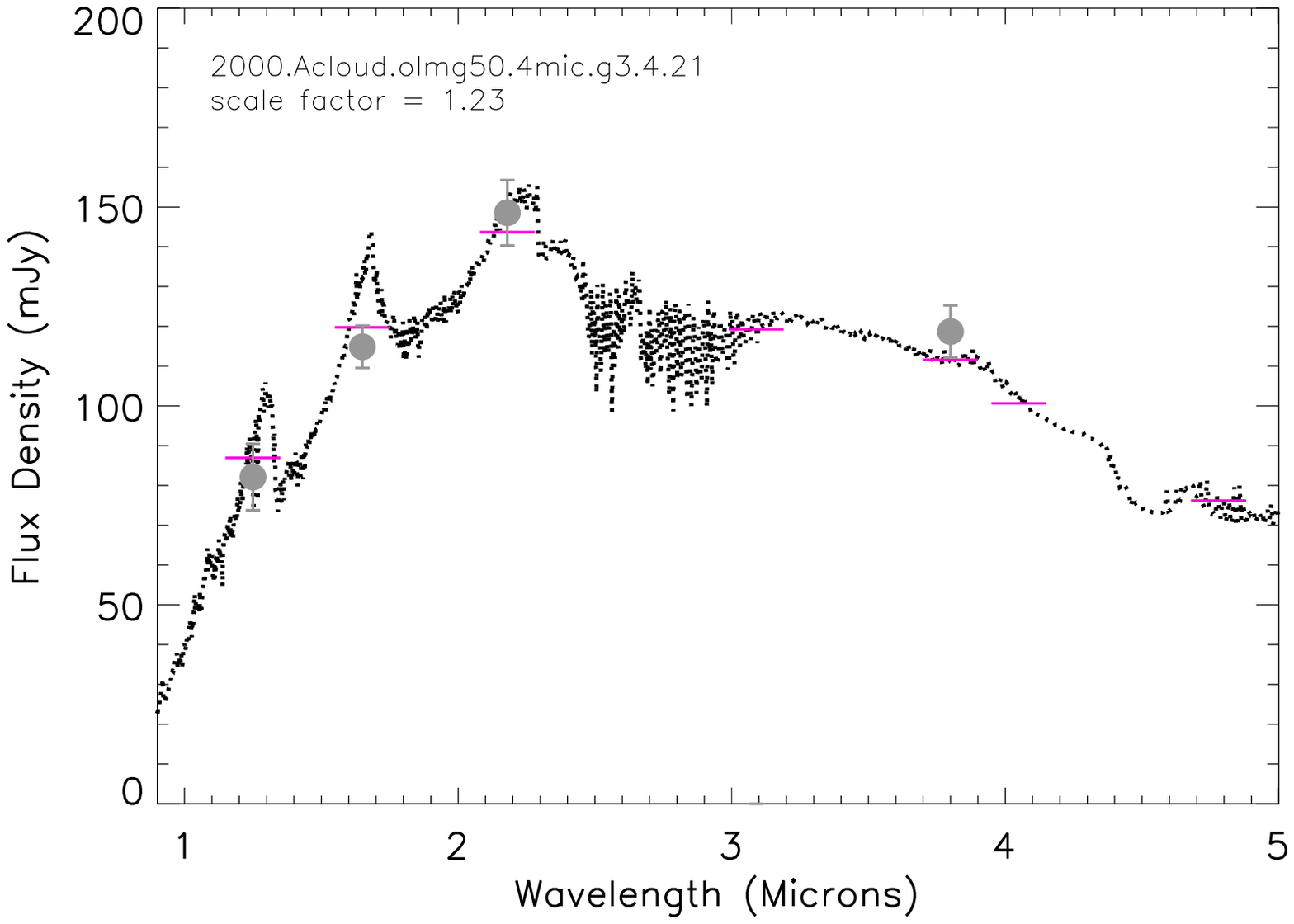}
\caption{Best-fit models fits to the photometry (grey dots) for each of the four atmospheric model grids (dark lines).  
The horizontal magenta lines show the predicted photometry from the models at the wavelengths with ROXs 42Bb detections plus longer wavelength [4.05] and $M^\prime$ filters studied in \citet{Currie2013a}.
}
\label{photfitbest}
\end{figure}

Table \ref{atmosfit} summarizes our model fitting results.  Figure \ref{photfit} displays the $\chi^{2}$ distributions 
for our photometric model fits while Figure \ref{photfitbest} shows the best-fit model spectrum and predicted photometry (magenta lines) 
with the photometric data overplotted.
The dust-free $COND$ models fail to reproduce the photometry (Figure \ref{photfit} and \ref{photfitbest}, top-left panels) 
at a statistically significant level 
($\chi^{2}_{min}$ $\sim$ 35.9 $>>$ $\chi^{2}_{95\%}$ $\sim$ 9.5).  The BT-Settl models (same figures, bottom-left panels) 
perform better but include only one model ($T_{eff}$ = 1800, log(g) = 3.5) consistent with the data at the 95\% confidence limit.  

Models incorporating substantial atmospheric dust and/or thick clouds best reproduce ROXs 42Bb's photometry.
The DUSTY models quantitatively yield the better fits than either COND or BT-Settl (top-right panel), with a $\chi^{2}$ minima of $\sim$ 2.47 
($T_{eff}$ = 1900 $K$, log(g) = 4) and models 
spanning $T_{eff}$ $\sim$ 1800--1900 $K$ consistent within the 95\% confidence limit.  The Burrows models reproduce the data 
even better (bottom-right panel) for $T_{eff}$ = 1800--2000 $K$ and log(g) = 3.6--4 and $T_{eff}$ = 1900--2000 $K$ with log(g) = 3.4
 ($\chi^{2}_{min}$ = 0.57).  While empirical comparisons to ROXs 42Bb's spectrum allowed temperatures of 1800--2600 $K$, 
modeling the object's photometry strongly favors the lower part of this range.

\subsection{Spectroscopy Fitting Results}
\begin{figure}
\centering
\includegraphics[scale=0.5]{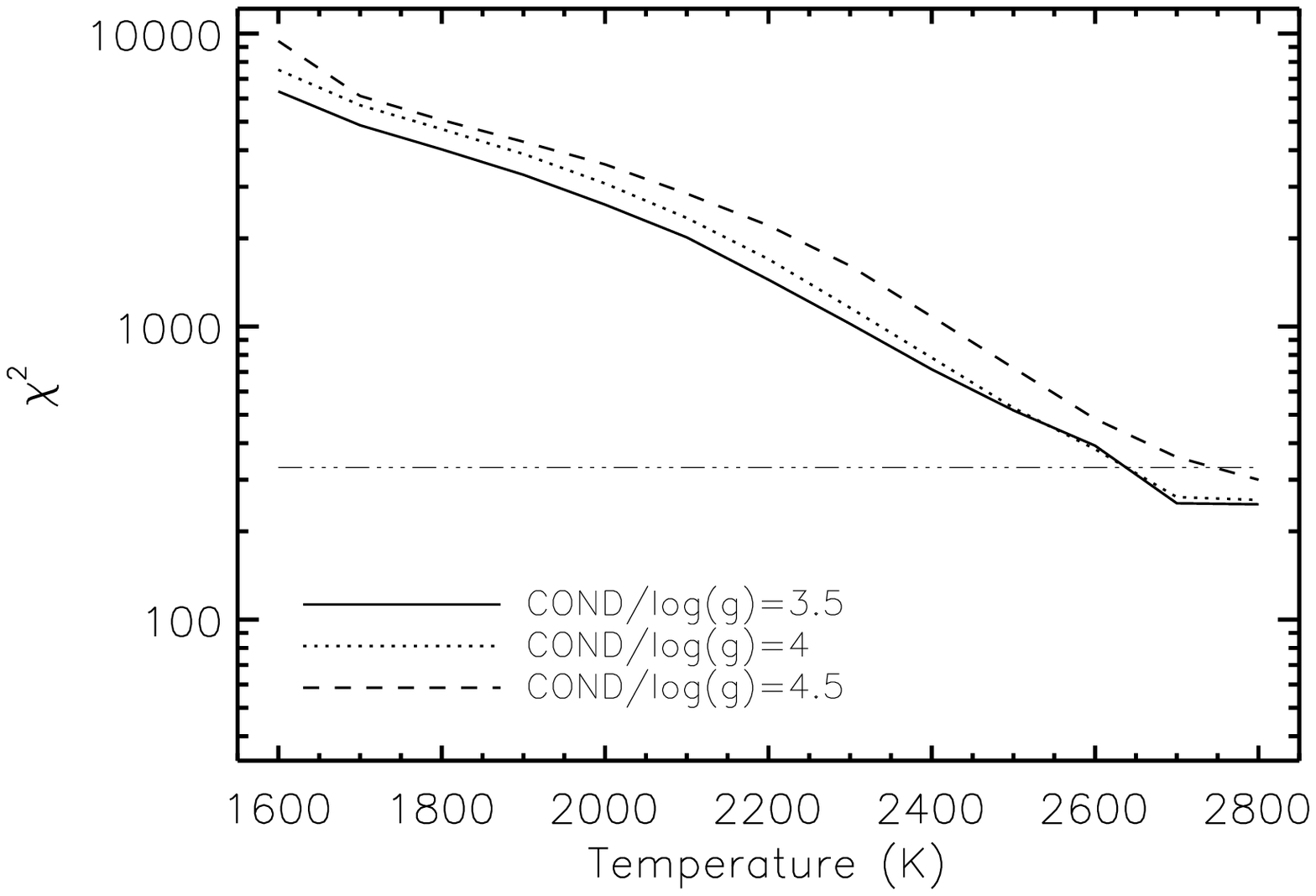}
\includegraphics[scale=0.5]{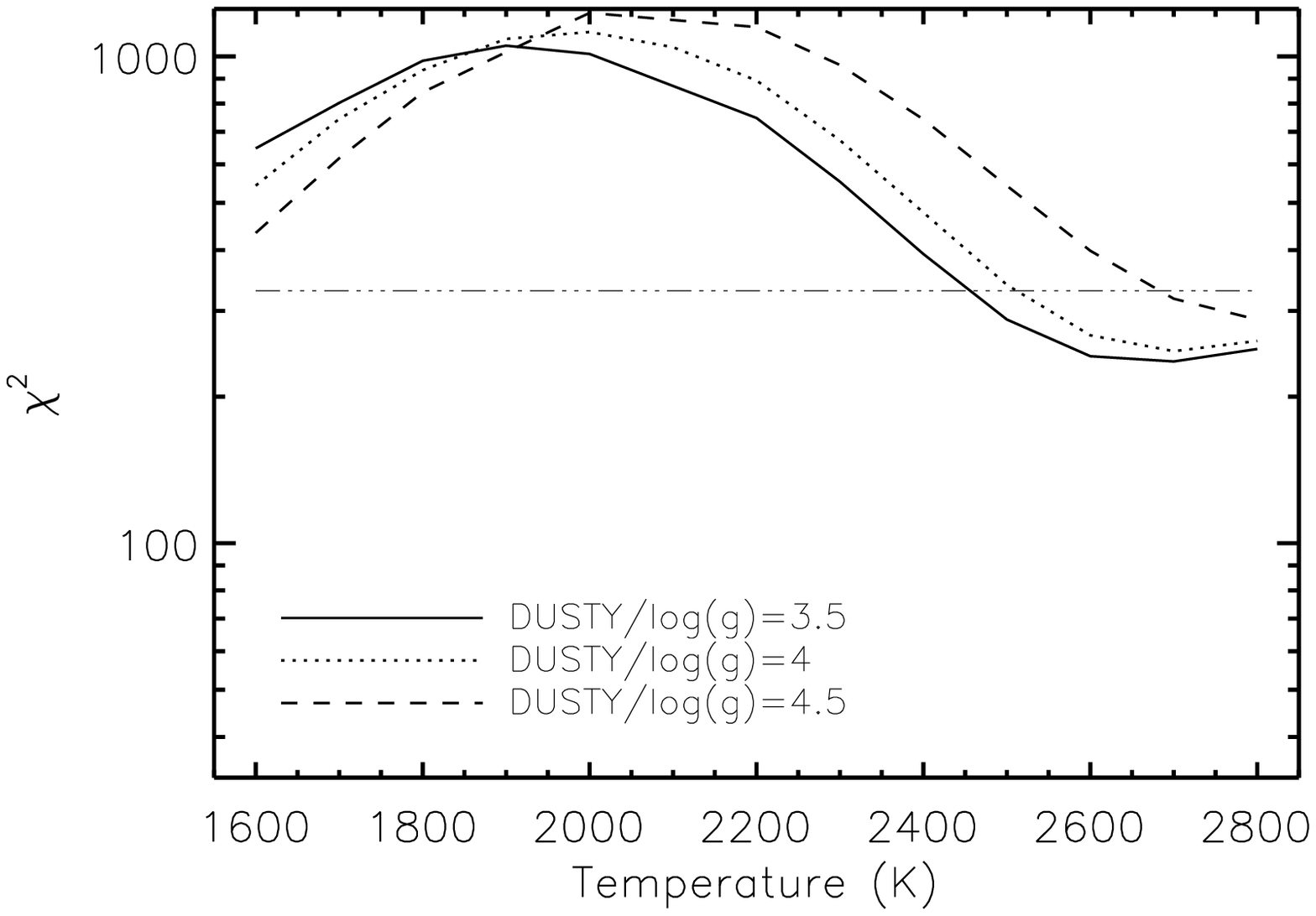}
\includegraphics[scale=0.5]{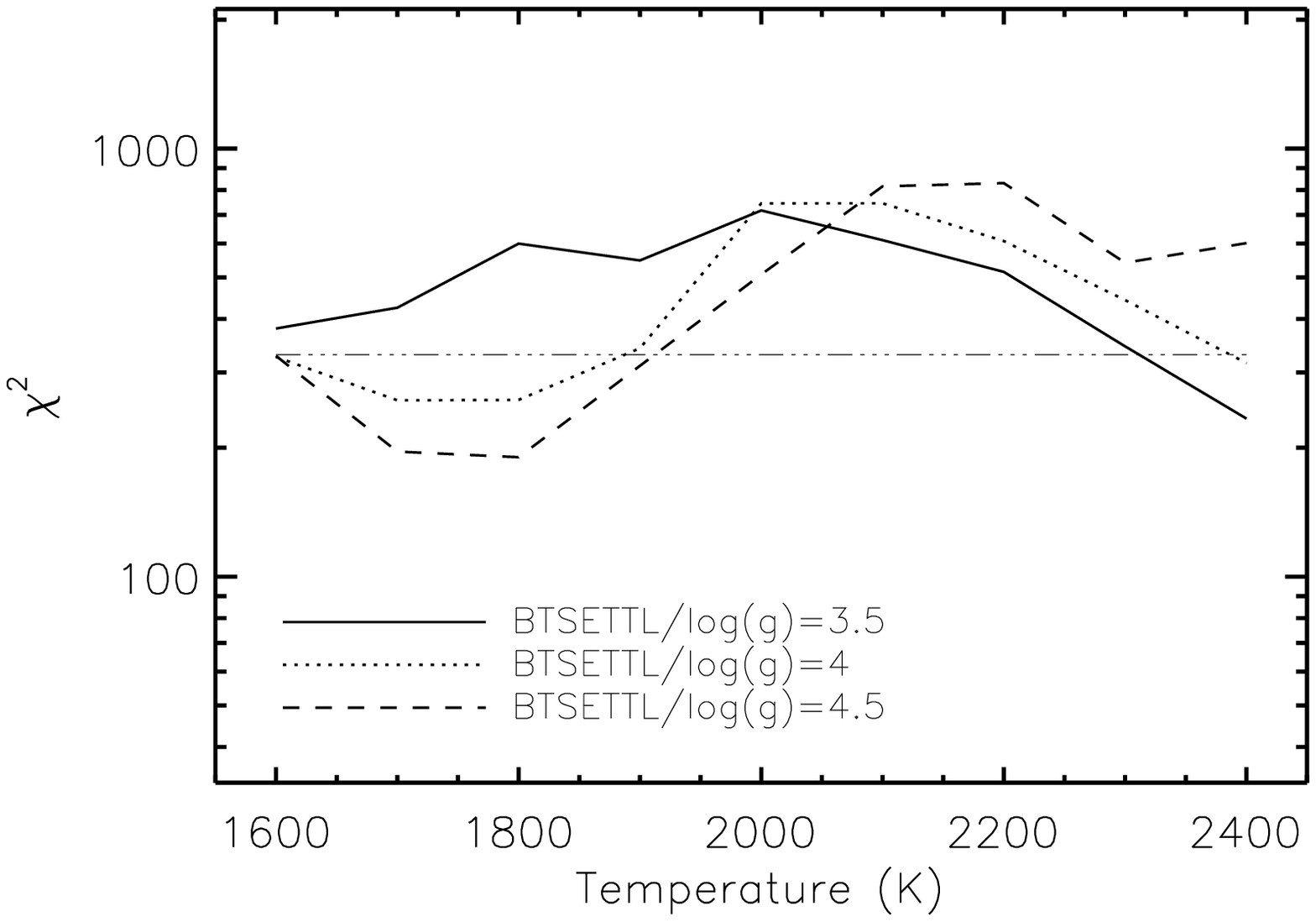}
\includegraphics[scale=0.5]{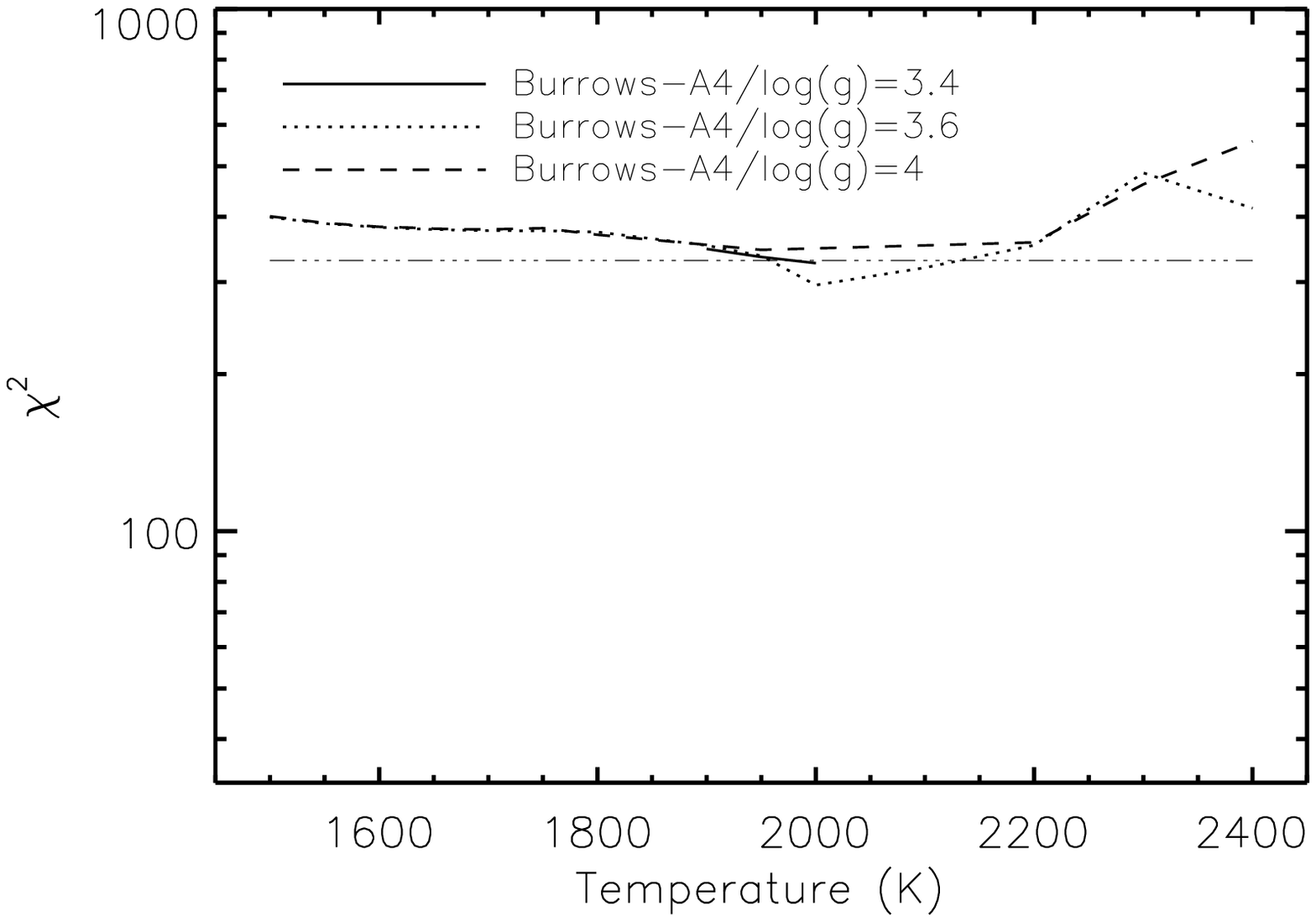}
\caption{Same as Figure \ref{photfit} except for the $K$-band spectrum.}
\label{specfit}
\end{figure}

\begin{figure}
\centering
\includegraphics[scale=0.5]{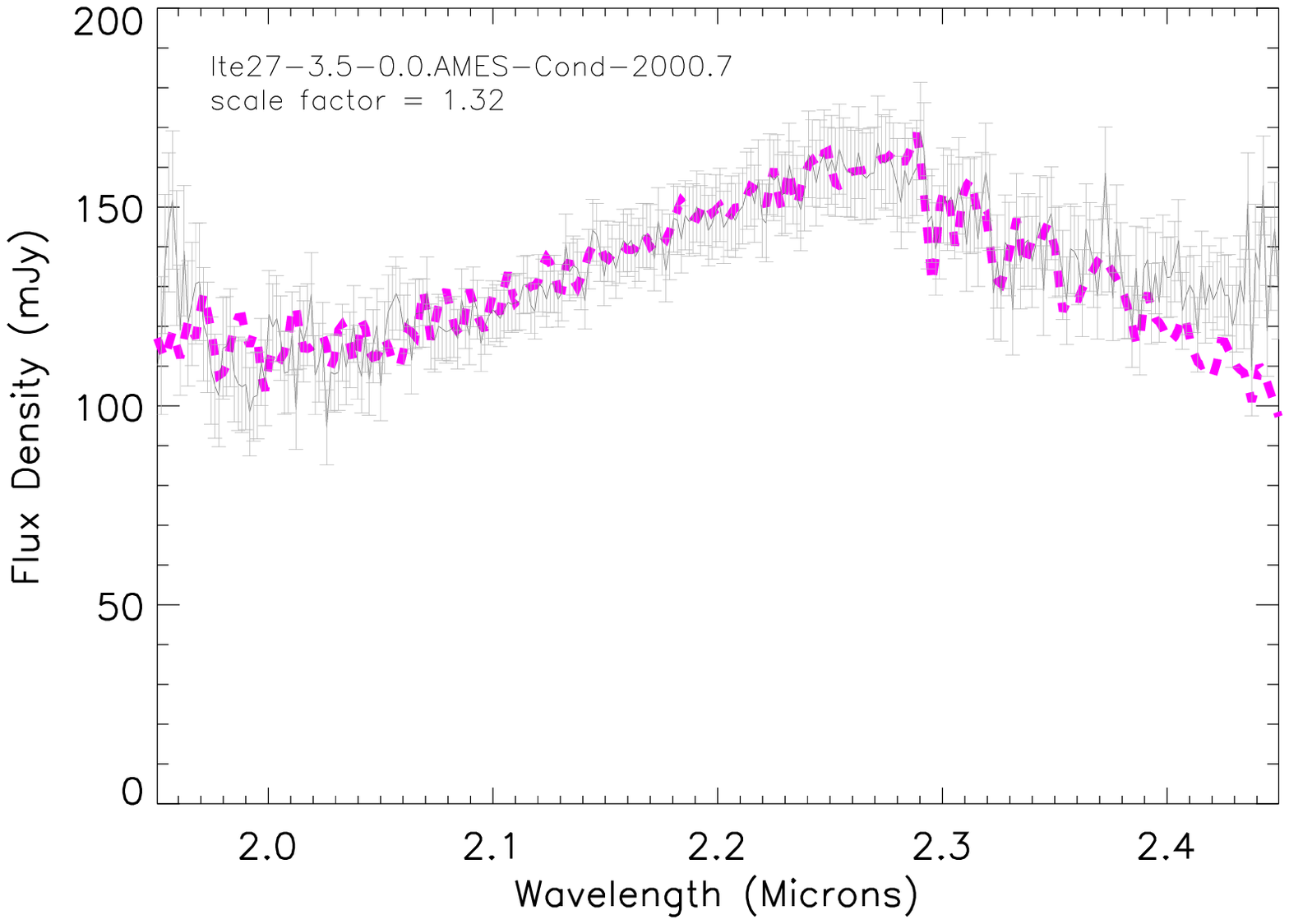}
\includegraphics[scale=0.5]{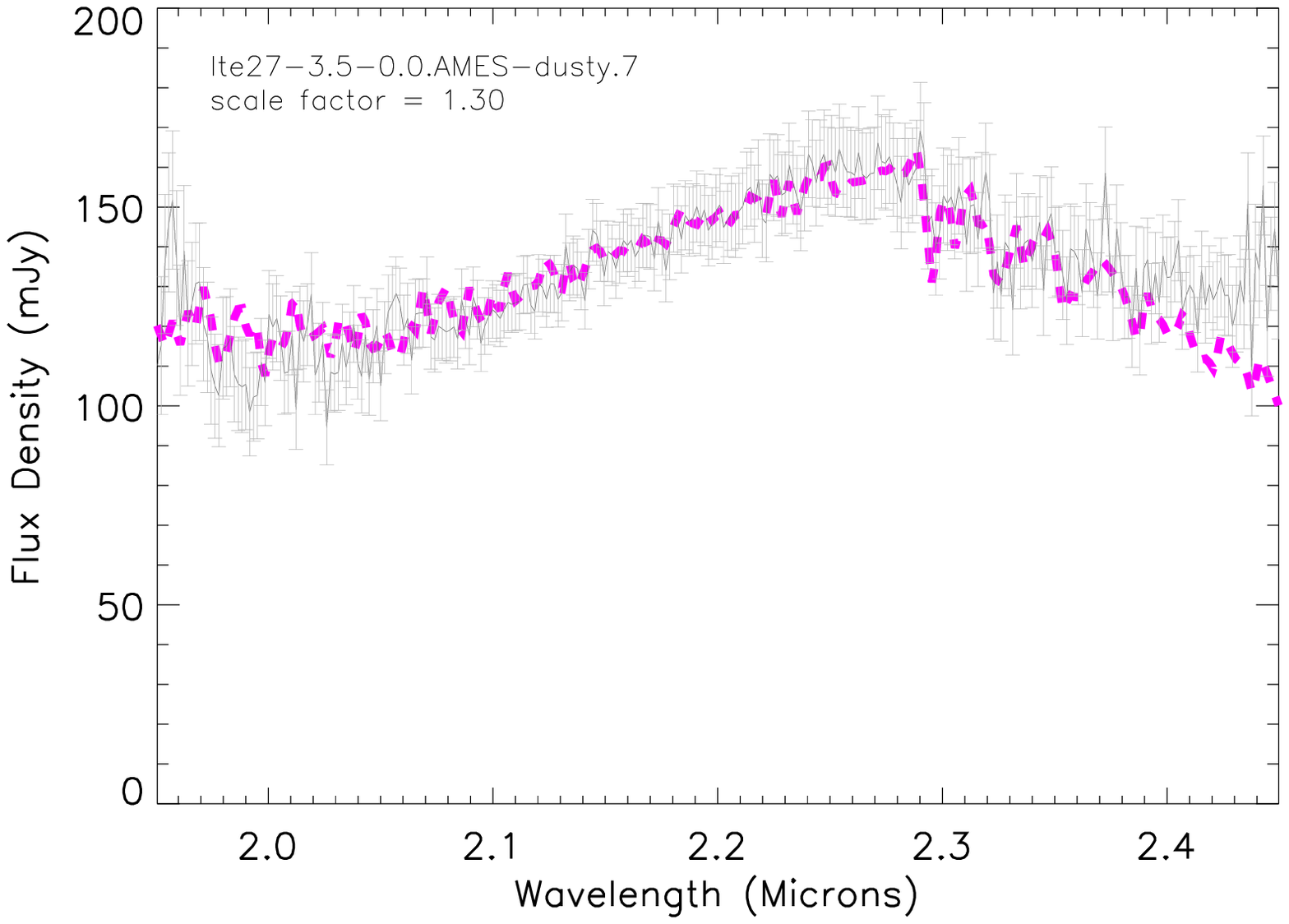}
\includegraphics[scale=0.5]{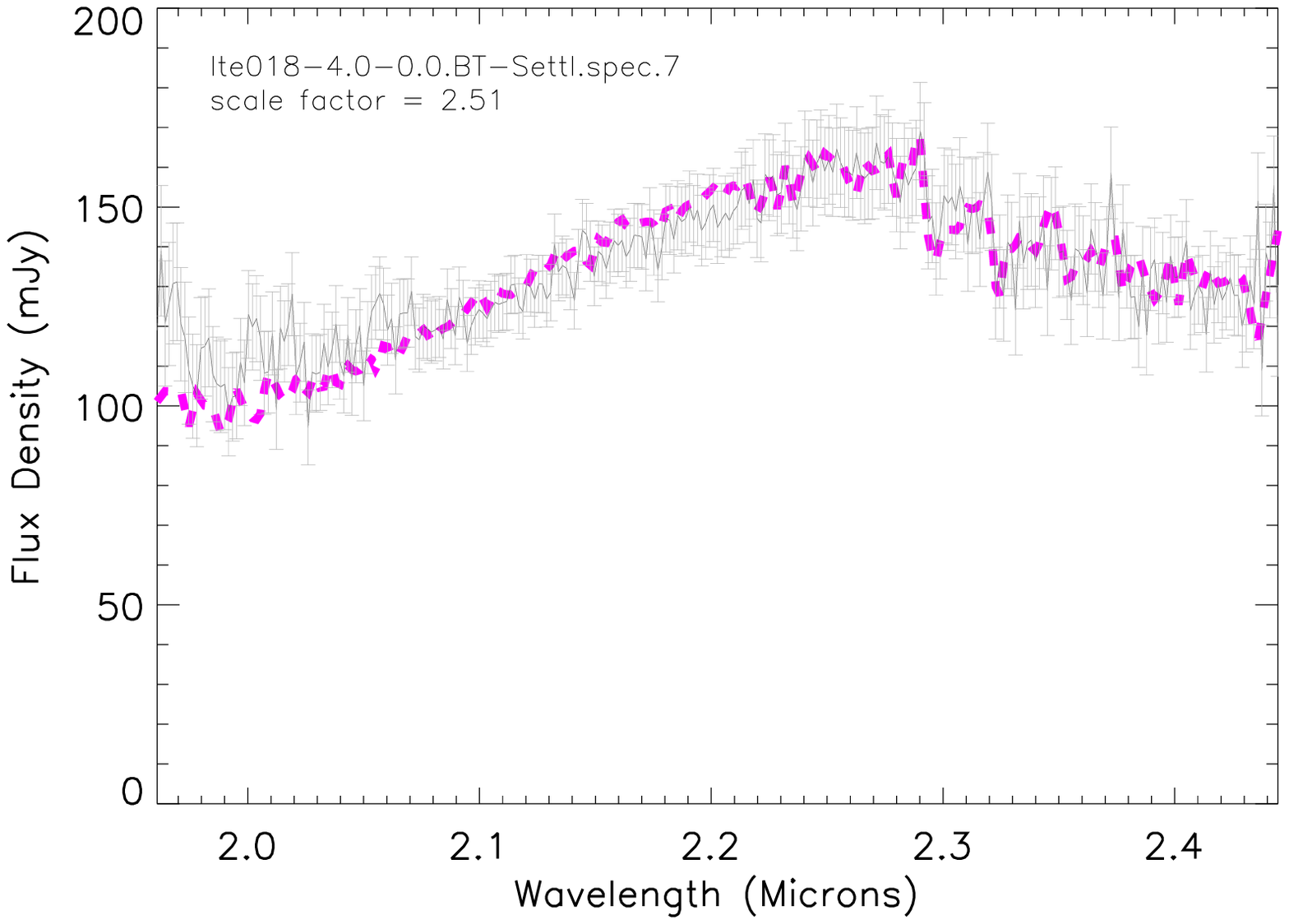}
\includegraphics[scale=0.5]{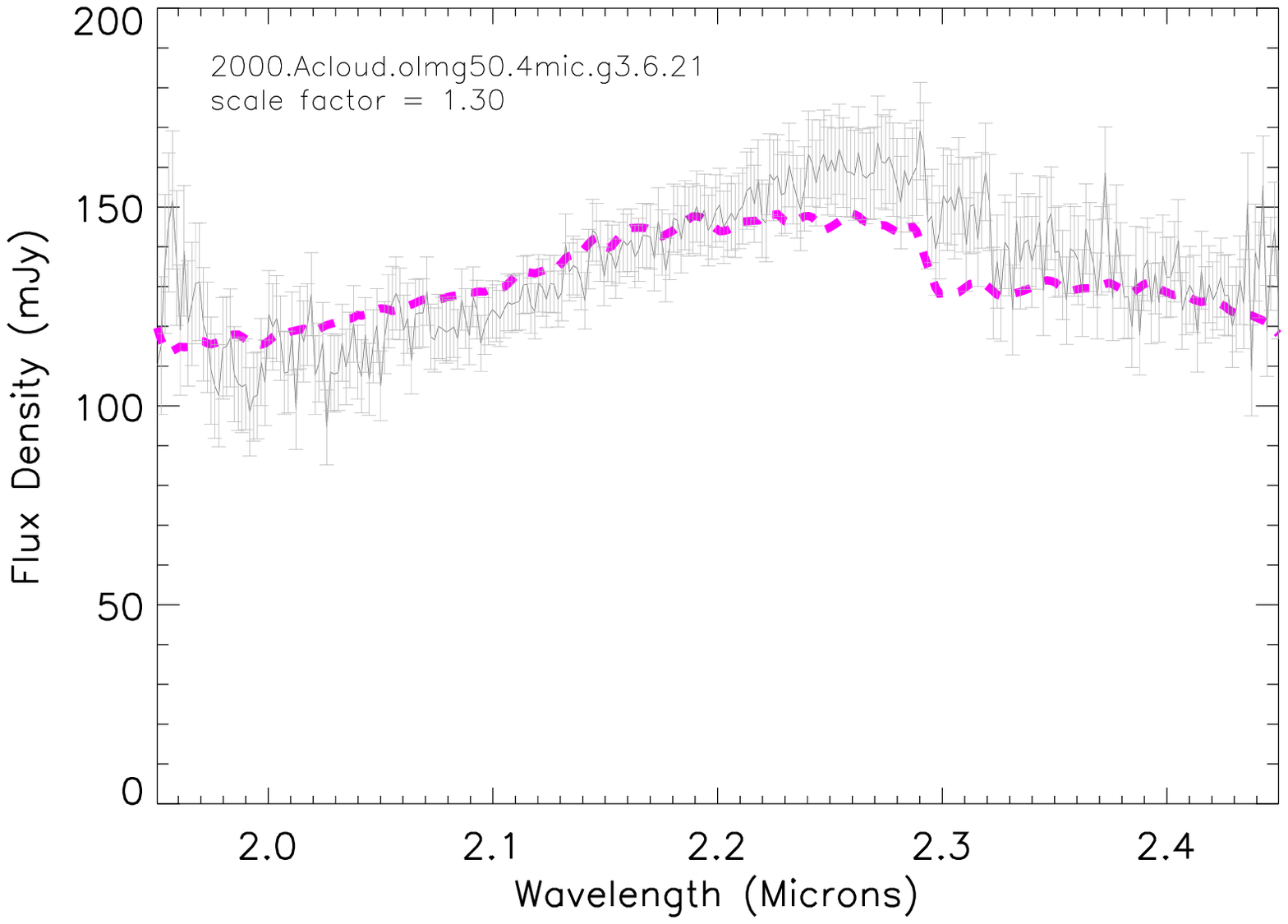}
\caption{Same as Figure \ref{photfitbest} except for the $K$-band spectrum.}
\label{specfitbest}
\end{figure}

Figures \ref{specfit} and \ref{specfitbest} display the $\chi^{2}$ distributions for our spectroscopic 
model fits and best-fit model with the $K$-band spectrum overplotted.  
The COND and DUSTY model spectra yield significant fits 
only at high temperatures ($T_{eff}$ = 2600--2800 $K$).  
 The BT-Settl models fit the spectrum well at 1700--1900 $K$/log(g) = 4--5 
and 2400 $K$/log(g) = 3.5, accurately reproducing the shape of most of the $K$-band peak and its location.
While the fit from the Burrows models tends to predict too flat of a $K$-band spectrum compared to BT-Settl, 
models with $T_{eff}$ = 1950--2100 $K$ and log(g) = 3.4--3.8 are consistent 
with the spectrum to within the 95\% confidence limit\footnote{While there is a slight preference for lower surface 
gravity models, the $\chi^{2}$ value much more strongly depends on temperature.}.  

\subsection{Derived Best-Fit Atmosphere Modeling Parameters}
\begin{figure}
\centering
\includegraphics[scale=0.5]{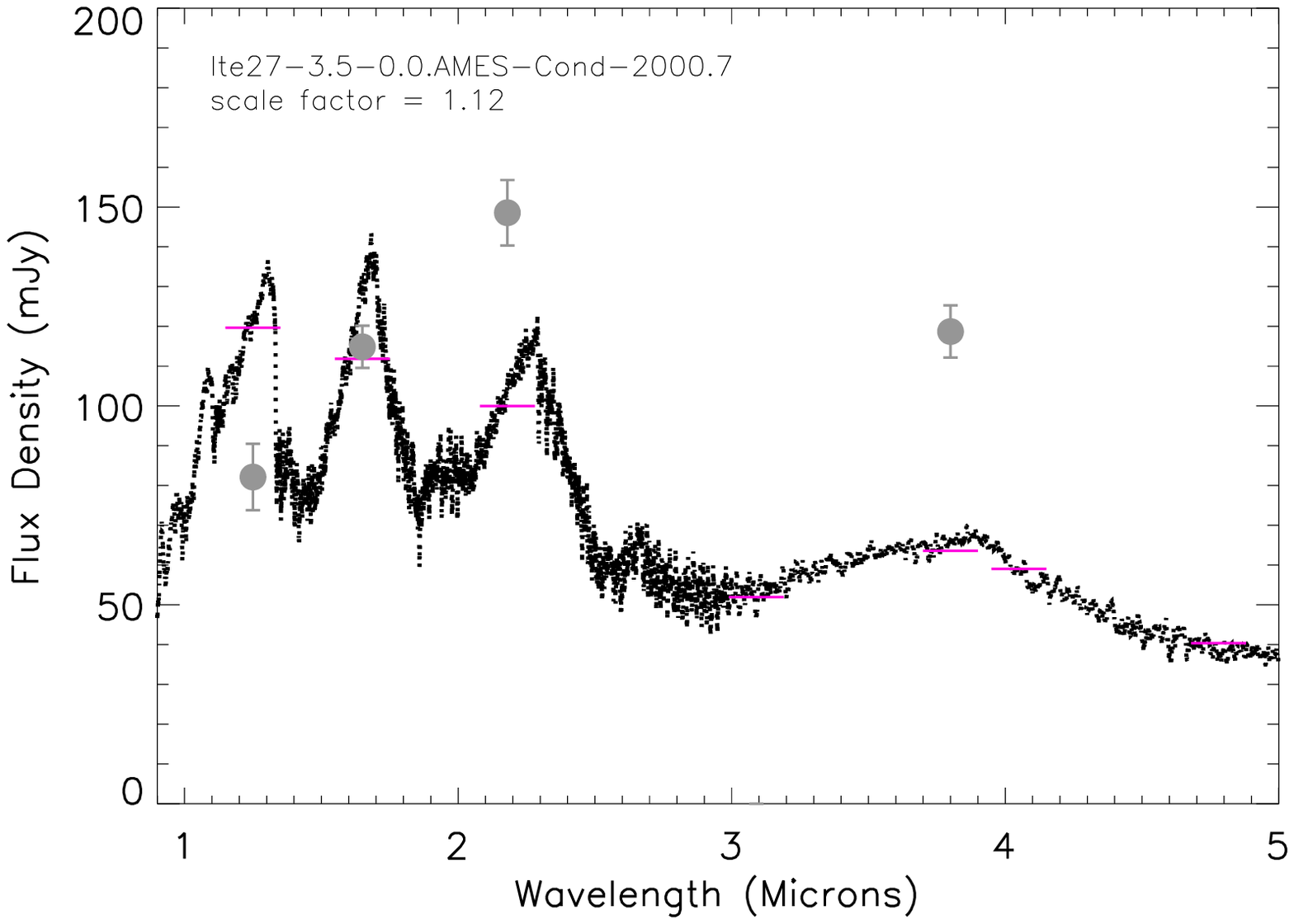}
\includegraphics[scale=0.5]{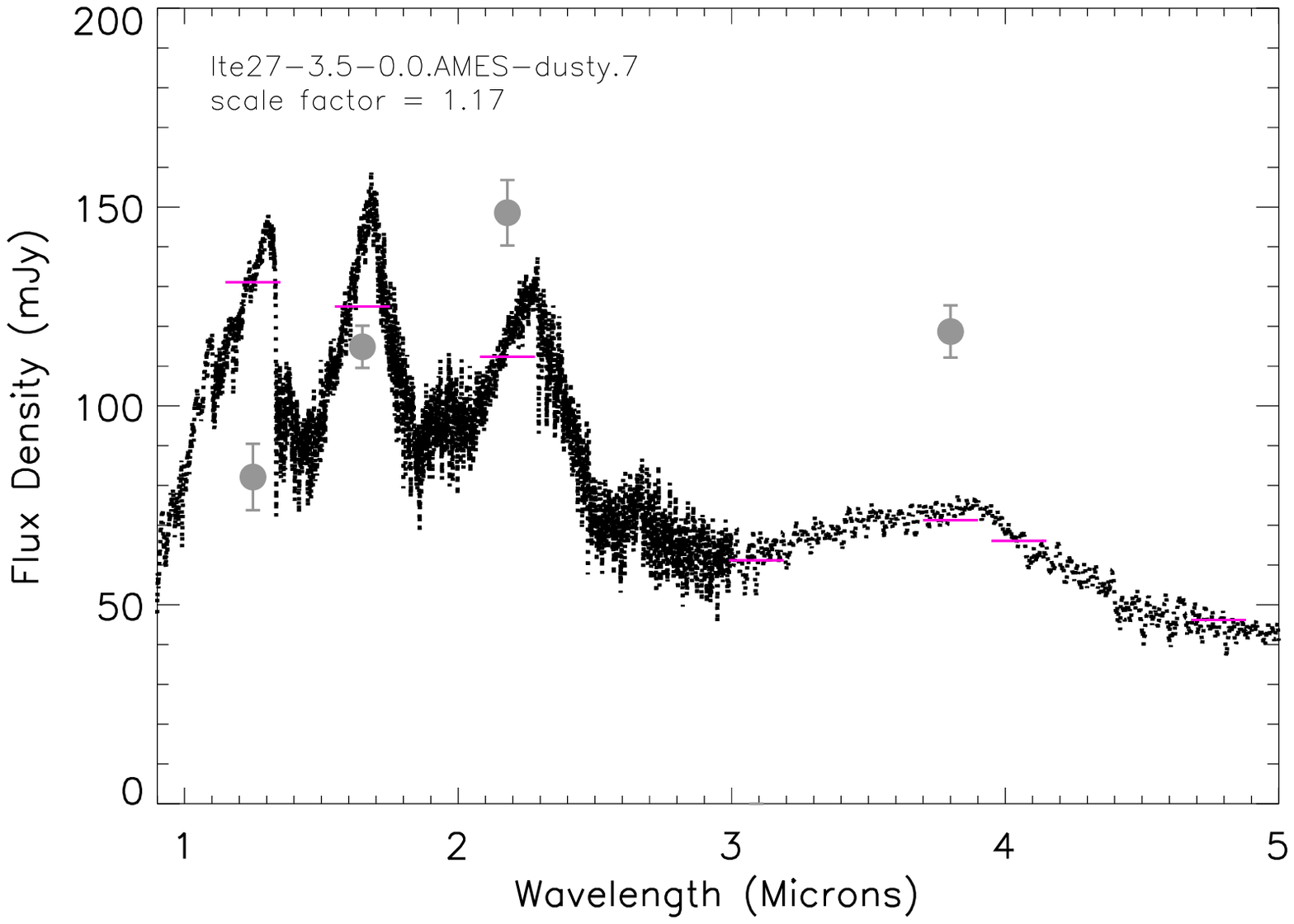}
\includegraphics[scale=0.5]{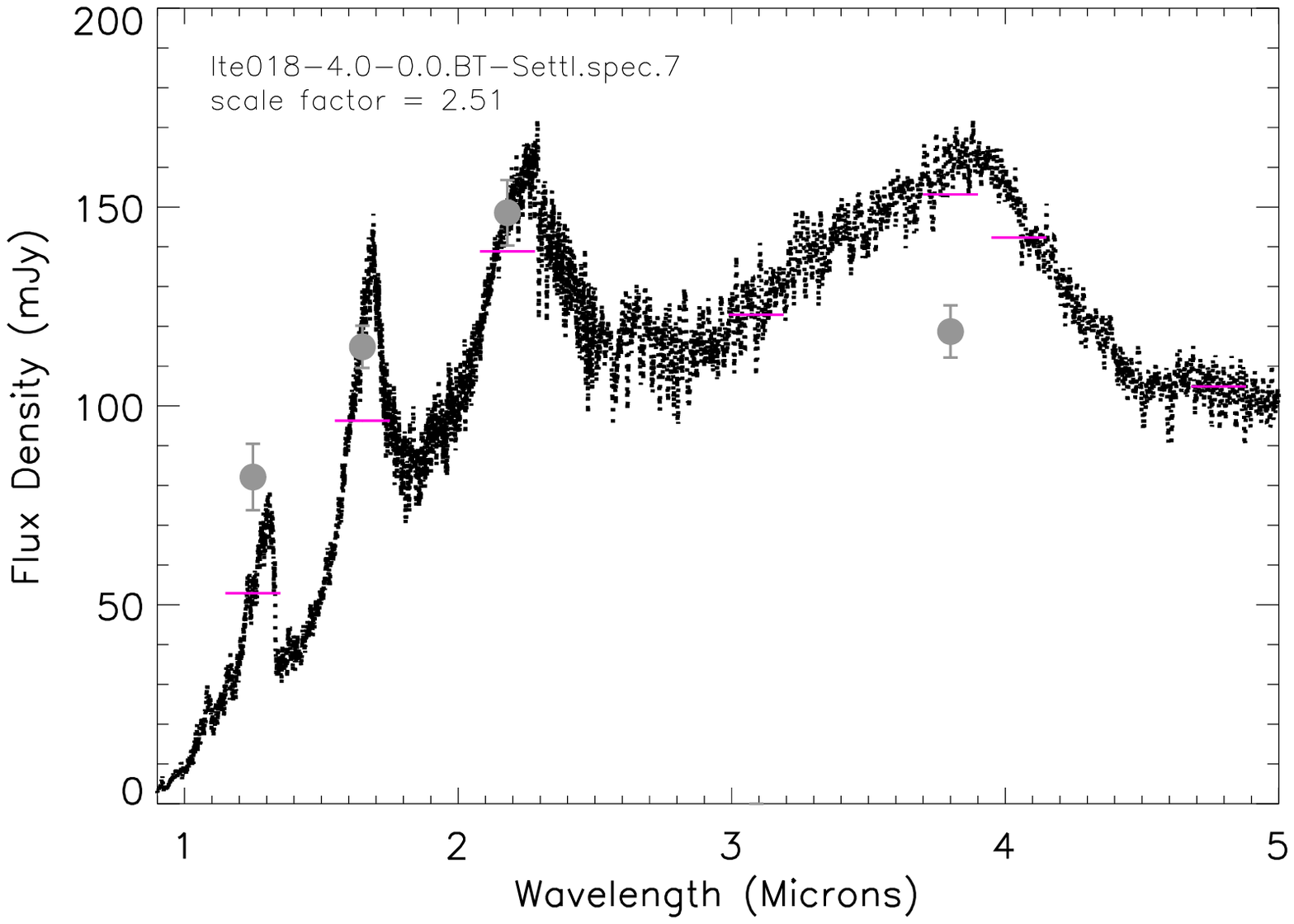}
\includegraphics[scale=0.5]{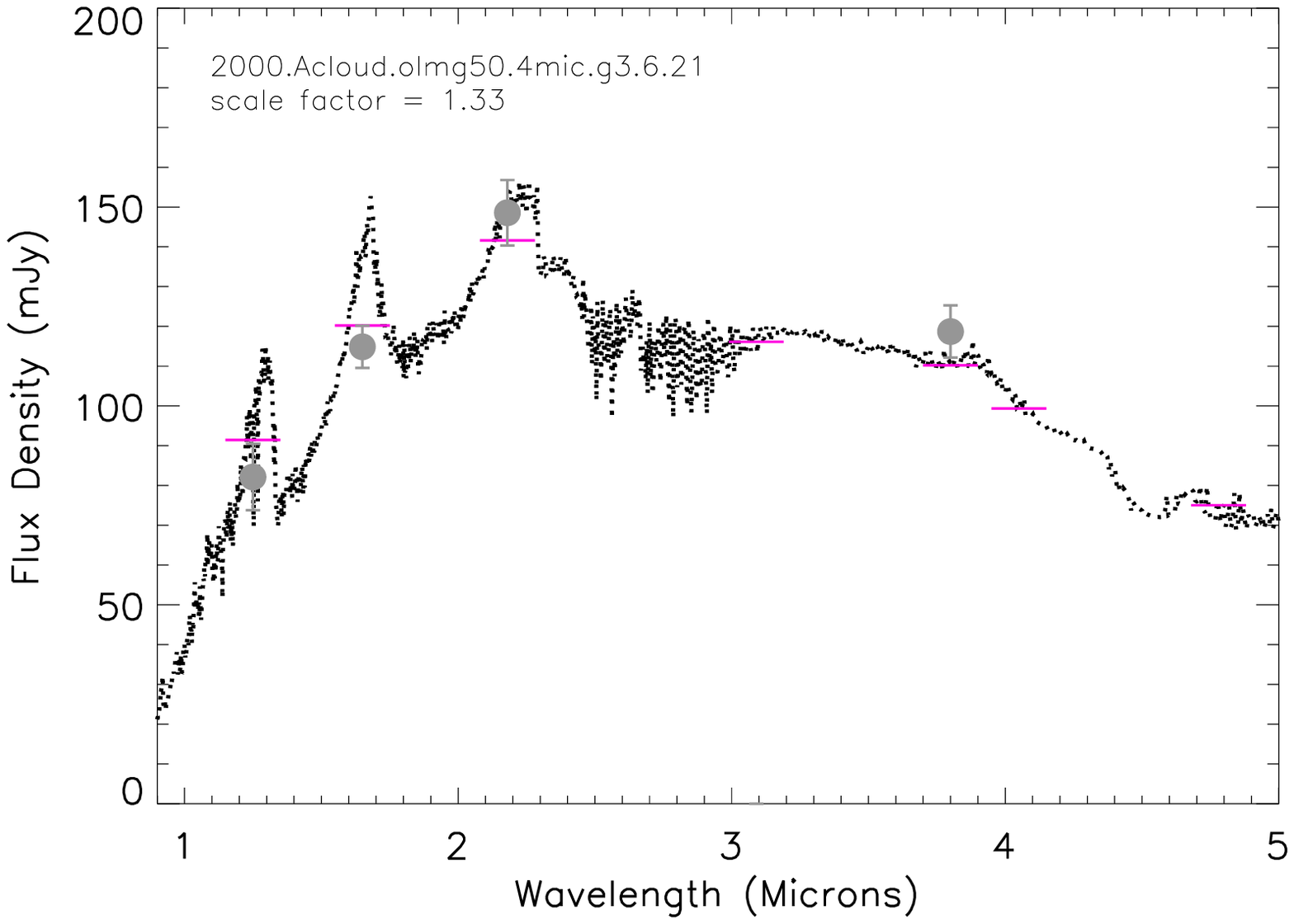}
\caption{Comparisons between models that best fit the ROXs 42Bb $K$-band spectrum and the ROXs 42Bb photometric 
data.  All $COND$ models, including those that accurately reproduce the $K$-band spectrum, fail to reproduce 
the companion's photometric data.  The $DUSTY$ and BT-Settl models that best fits the spectrum fail to match the companion's photometry.   
}
\label{photspeccomp}
\end{figure}

\begin{figure}
\centering
\includegraphics[scale=0.5]{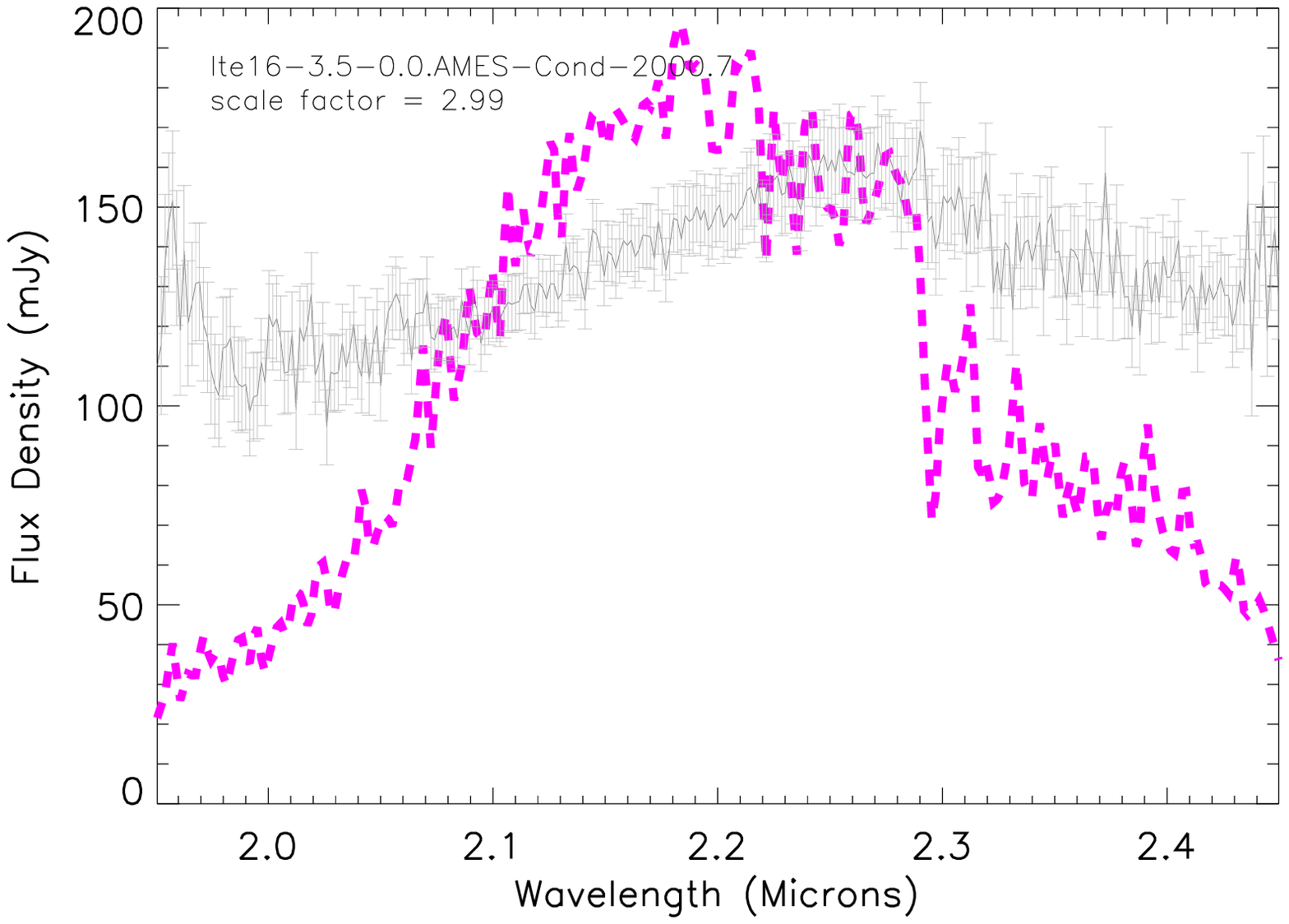}
\includegraphics[scale=0.5]{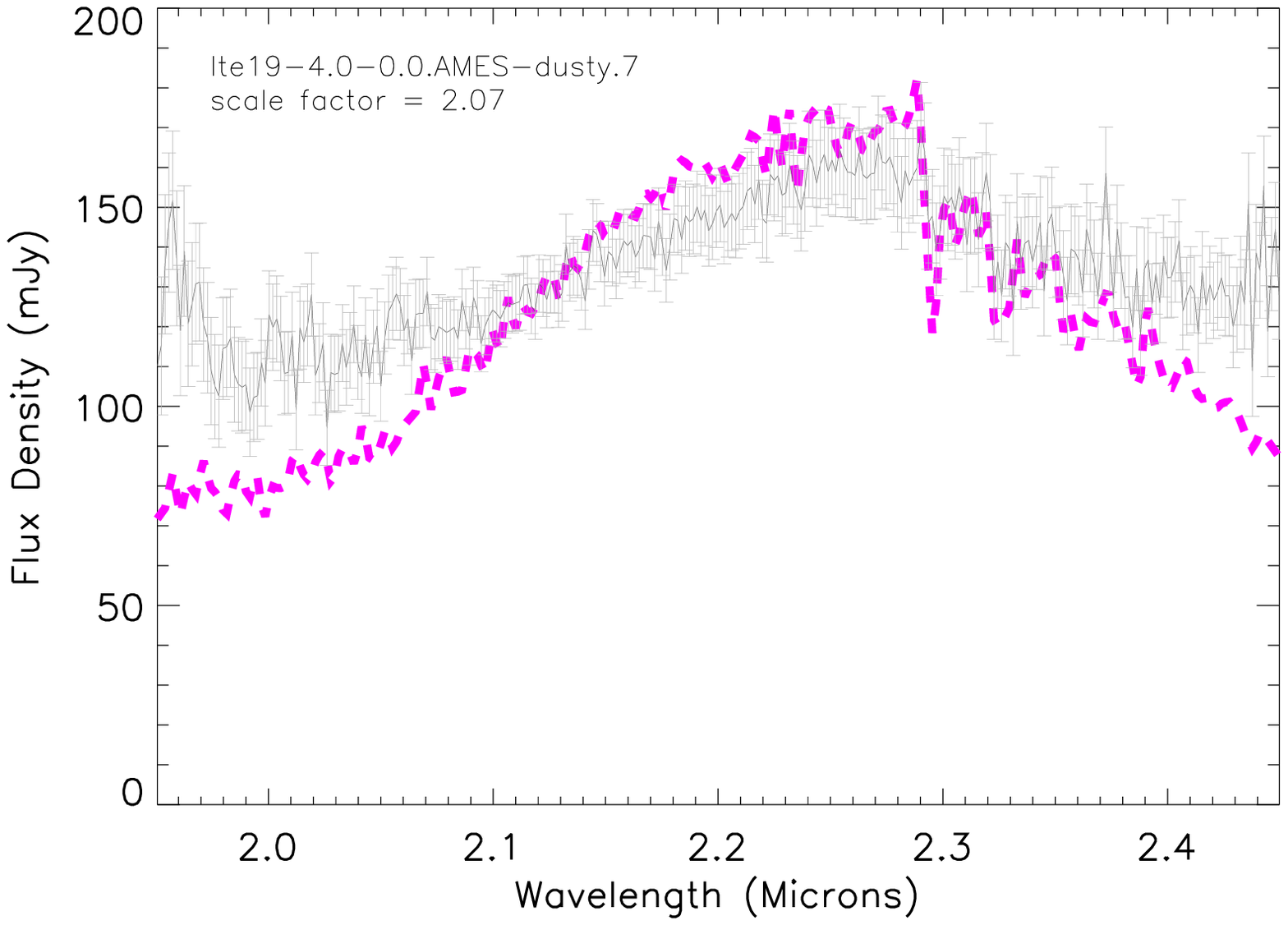}
\includegraphics[scale=0.5]{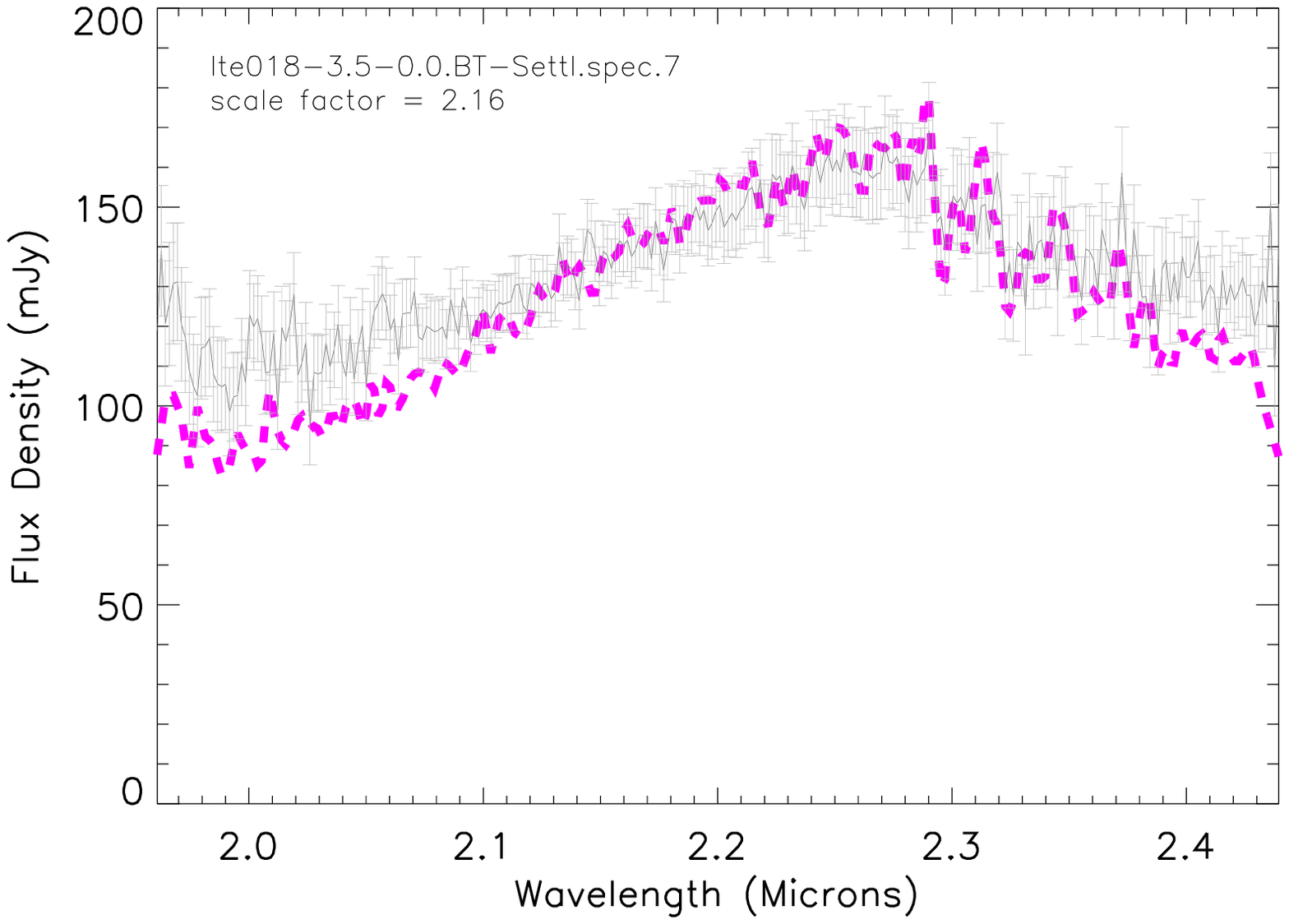}
\includegraphics[scale=0.5]{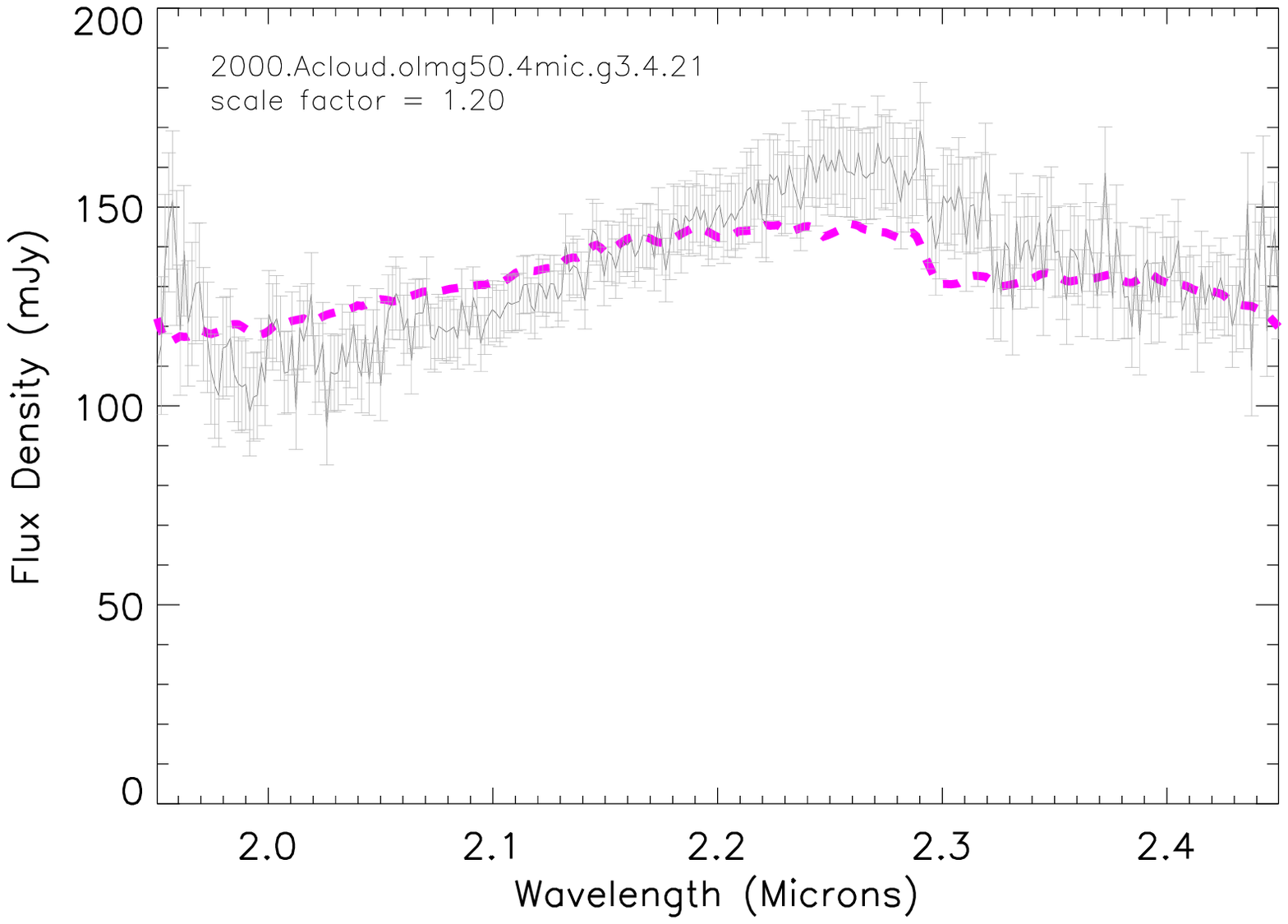}
\caption{Same as Figure \ref{photspeccomp} except comparing models that best fit the ROXs 42Bb \textit{photometry} to 
ROXs 42Bb's $K$-band spectrum.  The $COND$ and $DUSTY$ models clearly mismatch the $K$-band spectral shape.  
The BT-Settl model predicts slightly too narrow of a 2.2 $\mu m$ peak, while the Burrows model fares better.
A small range of parameter space -- $T_{eff}$ = 1950--2000 $K$, log(g) = 3.4--3.8 -- matches both the photometry and 
spectrum.}
\label{photspeccomp2}
\end{figure}

Figures \ref{photspeccomp} and \ref{photspeccomp2} and the righthand columns of Table \ref{atmosfit} clarify which model 
parameter space matches the ROXs 42Bb photometry and $K$-band spectrum simultaneously.
Although some $COND$ and $DUSTY$ models fit the $K$-band spectrum,
the same models fail to reproduce the ROXs 42Bb photometry.  
Similarly, the $COND$ and $DUSTY$ models best fitting the photometry fail to reproduce the $K$-band spectral shape.
 Thus, neither of the model limiting cases can simultaneously fit photometry
and spectra.  

The inferred temperatures for
BT-Settl models that best fit the spectrum and the photometry
are similar.  However, as with the COND and DUSTY models, the specific BT-Settl models yielding
fits to within the 95\% confidence limit for the spectra fail to match the photometry (Figure \ref{photspeccomp})
and vice-versa (Figure \ref{photspeccomp2}), owing to the model predicting too narrow of a 2.2 $\mu m$ peak.
In contrast to all other models, the Burrows models are able to yield some parameter space that fits \textit{both} the ROXs 42Bb photometry
and spectroscopy at the 95\% confidence limit.  For instance, the $T_{eff}$ = 2000 $K$, log(g) =3.6 model that best fits the spectrum also agrees with the 
photometric data to within the 95\% confidence limit (Figure \ref{photspeccomp}, bottom-right panel).

To derive best-estimated atmospheric parameters, we focus on the 
Burrows thick cloud models that simultaneously fit the photometry and $K$-band spectrum: 
$T_{eff}$ = 1950--2000 $K$, log(g) = 3.4--3.8.  Thus, spectral analysis (Sect 3.2) and atmospheric modeling (Sect. 4.1-4.2) 
combined shows ROXs 42Bb to have a very dusty/cloudy, low surface gravity L0 dwarf atmosphere with $T_{eff}$ = 1950--2000 $K$.  

\subsection{Inferred Properties from Best-Fitting Models}
The models matching ROXs 42Bb's photometry and spectroscopy allow us to derive an estimated 
radius for the object.  Compared to the nominal radii for a given $T_{eff}$ and 
log(g) assumed from the \citet{Burrows1997} evolutionary models, our fits require radii larger by 
factors of 1.25--1.53 for models with $T_{eff}$ = 1950 $K$ and log(g)=3.4--3.8.  For the $T_{eff}$ = 2000 $K$ and 
log(g) = 3.4--3.8 models, the scaling factors are smaller: 1.2--1.45.  In each case, the scaling factors that best fit 
 the photometry match those that best fit the spectrum to within $\sim$ 2.5\%.  We obtain best-fit radii of 
$R$/$R_{J}$ $\sim$ 2.55 $\pm$ 0.20 for the $T_{eff}$ = 1950 $K$ models and $R$/$R_{J}$ $\sim$ 2.43 $\pm$ 0.18 for 
the $T_{eff}$ = 2000 $K$ models, where the errors include consider the dispersion in radius scaling estimates 
amongst good-fitting models and the distance uncertainty of $\sim$ 5\%.

ROXs 42Bb's luminosity-estimated mass, derived from the temperature and radius of our fits, is consistent 
with the sub-deuterium burning, 9$^{+6}_{-3}$ $M_{J}$ estimate from \citet{Currie2014}.
  Luminosity estimates for the $T_{eff}$ = 1950 $K$ and 2000 $K$ models yield identical values: 
log(L/L$_{\odot}$) = -3.07 $\pm$ 0.07.  Given an estimate age of 2.5--3 $Myr$ for ROXs 42B \citep{Currie2014}, both the 
\citet{Baraffe2003} and \citet{Burrows1997} ``hot start" evolutionary models nominally predict a mass of 10$^{+2}_{-1}$ $M_{J}$.  
Displacing the evolutionary tracks by a characteristic protostar lifetime of $\tau_{Class 0/I}$ $\sim$ 0.5 $Myr$ \citep{Evans2009} 
lowers the masses to 9$^{+1.5}_{-1}$ $M_{J}$.  If instead ROXs 42B is 1 $Myr$ old like much of the $\rho$ Oph complex, 
ROXs 42Bb's mass is $\approx$ 6 $M_{J}$.  

Considering these uncertainties, we estimate a mass of 9$^{+3}_{-3}$ $M_{J}$ for ROXs 42Bb. 
We derive consistent (slightly lower, much higher) mass estimates if we take the surface gravity and radii of our best-fit 
models at face value: for log(g) = 3.6 (3.4, 3.8) we derive $M$ $\approx$ 9.1--10.4 $M_{J}$ (6--7.3 $M_{J}$, 14.3--15.8 $M_{J}$).

\section{Discussion}
Our analysis of ROXs 42Bb's 1--4 $\mu m$ photometry and $K$-band spectrum provides strong evidence that its 
atmosphere is dusty, with an effective temperature of $\sim$ 2000 $K$.  ROXs 42Bb's luminosity of log(L/L$_{\odot}$) 
$\sim$ -3.07 $\pm$ 0.07, when compared to evolutionary models, implies a mass of 6--12 $M_{J}$.  Thus, our atmosphere 
modeling supports earlier, largely empirically-driven arguments from \citet{Currie2014} 
that ROXs 42Bb is a sub-deuterium burning, planet-mass object.

Despite our success in identifying some acceptably fitting models as determined by a $\chi^{2}$ statistic, 
ROXs 42Bb's atmosphere still presents significant modeling challenges.  In particular, no model yet reproduces 
both the photometry and the triangular-shaped pseudo-continuum at $K$-band peaked at 2.24 $\mu m$.  
The primary mismatches between the best-fit Burrows models and the spectrum are that the models 
predict too flat of a spectrum, resulting in an overpredicted flux density at 2--2.1 $\mu m$ and underpredicted 
flux density over the 2.25--2.3 $\mu m$ region.  
As $H_{2}O$ absorption controls the shape of the spectrum at 2--2.15 $\mu m$ and CIA $H_{2}$ controls the slope of the spectrum at 
$\sim$ 2.2--2.3 $\mu m$, different treatments of these opacity sources may improve fits to the data\footnote{Note that 
the Burrows models are not alone in being strained to reproduce some gravity sensitive features in the spectra of 
young substellar/planet-mass objects.  For instance, the $H_{2}$(K) index appears to be a much stronger function of 
gravity than predicted from the Saumon and Marley models \citep[see][]{Canty2013}.}.

Our explored model parameter space is limited in important ways that may slightly modify the 
best inferred atmospheric properties.  For instance, as our model fits are only weakly sensitive to gravity, 
we adopted a narrow range of log(g) = 3.4--4.  Best-fit atmosphere models typically required larger radii by 20--30\% 
compared to their nominal values, which are tied to the \citet{Burrows1997} planet/brown dwarf luminosity 
evolution models, where the rescaling was smaller for lower gravity.  Exploring even lower gravities (e.g. log(g) = 3-3.2) 
may yield as good or better fits to the data without rescaling any radii and thus may be preferred.
Models with non-solar carbon abundances may provide marginally better fits to the photometry and/or 
spectra of hot, young planet-mass 
objects \citep{Barman2011,Galicher2011,Currie2013a}.  These issues should be considered to better constrain 
ROXs 42Bb's atmospheric properties, in particular its gravity and chemistry.

Finally, incorporating new near-IR spectra may further clarify ROXs 42Bb's 
temperature, gravity, and chemistry.  The $H_{2}O$ index at the blue edge of $H$ band \citep{Allers2013} provides a 
gravity insensitive independent check on our $K$-band derived spectral type as does the $H_{2}O-1$ index near $J$ 
band \citep{Slesnick2004}.   The depth of $K I$ lines in $J$ band spectra of substellar objects is gravity sensitive 
\citep{Allers2007,Luhman2007} and provides another way to compare ROXs 42Bb's atmosphere with those for other 
young substellar objects, especially those in Taurus and Upper Scorpius\footnote{While a detailed 
comparison with \citet{Bowler2014}'s $J$ and $H$ spectra is beyond the scope of this paper, 
they did provide some evidence that ROXs 42Bb's spectrum at $J$ and $H$ band bears strong resemblance to 
young planet-mass objects.  Our best-fit atmosphere models do appear to reproduce key features of ROXs 42Bb 
at these wavelengths: the depth of the $K I$ lines at $J$, the sharp drop at 1.3 $\mu m$, and the triangular 
shape of the $H$ band spectrum.}.



\acknowledgements 
We thank Thorsten Ratzka, Ernst De Mooj, and Scott Kenyon for helpful draft comments and 
the anonymous referee for suggestions improving the presentation and data analyses in our paper.
This research has made use of the Keck Observatory Archive (KOA), which is operated by the W. M. Keck
Observatory and the NASA Exoplanet Science Institute (NExScI), under
contract with the National Aeronautics and Space Administration.  
TC and SD are supported by McLean Postdoctoral Fellowships.

{}

\begin{deluxetable}{lllllllll}
\tablecolumns{9}
\tabletypesize{\tiny}
\tablecaption{Near-to-mid Infrared Photometry}
\tablehead{{Object}&{Filter}& $\Delta$(mag) & 
Apparent Magnitude & Dereddened Absolute Magnitude}
\startdata
ROXs 42Bb & $J$ & 7.00 $\pm$ 0.11 & 16.91 $\pm$ 0.11 & 10.72 $\pm$ 0.17\\
          & $H$ & 6.86 $\pm$ 0.05 & 15.88 $\pm$ 0.05 & 9.87 $\pm$ 0.14\\
          & $K_{s}$  &6.34 $\pm$ 0.06 & 15.01 $\pm$ 0.06 & 9.13 $\pm$ 0.14\\  
          & $L^\prime$ &5.64 $\pm$ 0.06 & 14.06 $\pm$ 0.06 & 8.30 $\pm$ 0.15\\
"ROXs 42Bc" & $J$ & 6.69 $\pm$ 0.12 & 16.60 $\pm$ 0.12 & -\\
          & $H$ & 6.94 $\pm$ 0.09 & 15.96 $\pm$ 0.09 & -\\
          & $K_{s}$ &6.78 $\pm$ 0.07 & 15.45 $\pm$ 0.07 & -\\  
          & $L^\prime$ &6.76 $\pm$ 0.20 & 15.18 $\pm$ 0.20 & -\\
 \enddata
\tablecomments{New data is from Keck PID HD242N2 (P.I. A. Kraus).  Assuming a reddening of $A_{V}$ = 1.9 and an 
intrinsic color of $K_{s}$-$L^\prime$ = 0.14 $\pm$ 0.05 for ROXs 42B, we estimate an 
apparent $L^\prime$ magnitude of 8.42 $\pm$ 0.05.  The ROXs 42Bb $H$ band photometry 
was previously reported in \citet{Currie2014}.}
\label{newphot}
\end{deluxetable}

\begin{deluxetable}{lcllccccccc}
\setlength{\tabcolsep}{0pt}
\tablecolumns{7}
\tablecaption{Atmosphere Modeling Grid}
\tiny
\tablehead{{\textbf{Model}}&{}&{\textbf{Range}}&{}&{}\\
{}&{$T_{eff}$ ($K$)}&{$\Delta$$T_{eff}$}&{log(g)} & {$\Delta$(log(g))}}
\startdata
\textit{Limiting Cases}\\
AMES-COND & 1600-2800 & 100 & 3.5--5& 0.5 \\
AMES-DUSTY  &1600-2800& 100 & 3.5--5& 0.5 \\
\textit{Cloud Models}\\
BT-Settl & 1500-2400 & 100 & 3.5--5& 0.5\\
Burrows/A4 & 1500-1800,2100-2400 & 100 & 3.6--4 &0.2\\
   & 1900-2000 & 50 & 3.4--4 &0.2 \\
 \enddata
\label{atmosparam}
\end{deluxetable}

\begin{deluxetable}{llcccc}
 \tiny
\tabletypesize{\tiny}
\tabletypesize{\small}
\tablecolumns{11}
\tablecaption{Model Fitting Results}
\tablehead{{}&{  \textbf{Photometry}}&{  \textbf{Spectroscopy}}\\
{Model}& {T$_{eff}$ (K), log(g) (95\%)} & {T$_{eff}$ (K), log(g) (95\%)}}
\startdata
\textit{Limiting Cases}\\
AMES-COND &- & 2600-2800, 3.5--4 \\
          &- & 2800, 4.5\\
AMES-DUSTY &  1800--1900, 3.5--5 & 2500--2800, 3.5--4 \\
           &  1900, 4            & 2700--2800, 4.5\\
           &                     & 2700, 5\\
\\
\textit{Cloud Models}\\
BT-Settl      & 1800, 3.5 & 1700-1900, 4--5 & -- \\
              &           & 2400, 3.5--4 &--\\
Burrows/A4    & 1800--2000, 3.6--4.25 & 1950-2100, 3.4--3.8\\
              & 1900--2000, 3.4       &        \\
\hline
\hline
\textit{Models Fitting Photometry \textit{and} Spectra}\\
Burrows/A4    & 1950--2000, 3.4--3.8
 \enddata
\label{atmosfit}
\end{deluxetable}



\begin{thebibliography}{}
\bibitem[Allard et al.(2001)]{Allard2001}Allard, F., et al., 2001, \apj, 556, 357
\bibitem[Allard et al.(2012)]{Allard2012}Allard, F., et al., 2012, in EAS Publications Series, Vol. 57. Ed. 
C. Reyle, C. Charbonnel, \& M. Schultheis, 3--43
\bibitem[Allers et al.(2007)]{Allers2007}Allers, K., et al., 2007, \apj, 657, 511
\bibitem[Allers and Liu(2013)]{Allers2013}Allers, K., Liu, M., 2013, \apj, 772, 79
\bibitem[Baraffe et al.(1998)]{Baraffe1998}Baraffe, I., et al., 1998, A\&A, 337, 403
\bibitem[Baraffe et al.(2003)]{Baraffe2003}Baraffe, I., et al., 2003, A\&A, 402, 701
\bibitem[Barman et al.(2011)]{Barman2011}Barman, T., et al., 2011, \apj, 735, L39
\bibitem[Bejar et al.(2008)]{Bejar2008}Bejar, V., et al., 2008, \apj, 673, L185
\bibitem[Bonnefoy et al.(2013)]{Bonnefoy2013}Bonnefoy, M., et al., 2013, A\&A in press\\
\bibitem[Bowler et al.(2011)]{Bowler2011}Bowler, B., et al., 2011, \apj, 743, 148
\bibitem[Bowler et al.(2014)]{Bowler2014}Bowler, B., et al., 2014, \apj\ in press
\bibitem[Burrows et al.(1997)]{Burrows1997}Burrows, A., et al., 1997, \apj, 491, 856
\bibitem[Burrows et al.(2006)]{Burrows2006}Burrows, A., et al., 2006, \apj, 640, 1063
\bibitem[Canty et al.(2013)]{Canty2013}Canty, J., et al., 2013, \mnras, 435, 2650
\bibitem[Cardelli et al.(1989)]{Cardelli1989}Cardelli, J., et al., 1989, \apj, 345, 245
\bibitem[Chauvin et al.(2004)]{Chauvin2004}Chauvin, G., et al., 2004, A\&A, 425, 29L
\bibitem[Currie et al.(2010)]{Currie2010b}Currie, T., Hernandez, J., et al., 2010, \apjs, 186, 191
\bibitem[Currie et al.(2011)]{Currie2011a}Currie, T., Burrows, A., et al., 2011, \apj, 729, 128
\bibitem[Currie et al.(2012a)]{Currie2012a}Currie, T., et al., 2012a, \apj, 760, L32
\bibitem[Currie et al.(2012b)]{Currie2012b}Currie, T., et al., 2012b, \apj, 755, L34
\bibitem[Currie et al.(2013)]{Currie2013a}Currie, T., et al., 2013, \apj, 776, 15
\bibitem[Currie et al.(2014)]{Currie2014}Currie, T., et al., 2014, \apj, 780, L30
\bibitem[Cushing et al.(2005)]{Cushing2005}Cushing, M., et al., 2005, \apj, 623, 1115
\bibitem[Evans et al.(2009)]{Evans2009}Evans, N., et al., 2009, \apjs, 181, 321
\bibitem[Galicher et al.(2011)]{Galicher2011}Galicher, R., et al., 2011, \apj, 739, L41
\bibitem[Ireland et al.(2011)]{Ireland2011}Ireland, M., et al., 2011, \apj, 726, 113
\bibitem[Kirkpatrick et al.(2006)]{Kirkpatrick2006}Kirkpatrick, D., et al., 2006, \apj, 639, 1120
\bibitem[Kraus et al.(2014)]{Kraus2014}Kraus, A., et al., 2014, \apj\ in press
\bibitem[Kuzuhara et al.(2013)]{Kuzuhara2013}Kuzuhara, M., et al., 2013, \apj, 774, 11
\bibitem[Lafreni\`ere et al.(2008)]{Lafreniere2008}Lafreni\`ere, D., et al., 2008, \apj, 689, 153L
\bibitem[Lafreni\`ere et al.(2010)]{Lafreniere2010}Lafreni\`ere, D., et al., 2010, \apj, 719, 497
\bibitem[Leggett et al.(2010)]{Leggett2010}Leggett, S., et al., 2010, \apj, 710, 1627
\bibitem[Lodieu et al.(2008)]{Lodieu2008}Lodieu, N., et al., 2008, \mnras, 383, 1385
\bibitem[Luhman et al.(2004)]{Luhman2004}Luhman, K., Peterson, D., Megeath, S. T., 2004, \apj, 617, 565
\bibitem[Luhman et al.(2007)]{Luhman2007}Luhman, K., et al., 2007, \apj, 659, 1629
\bibitem[Mamajek(2008)]{Mamajek2008}Mamajek, E., 2008, AN, 329, 10
\bibitem[Marois et al.(2006)]{Marois2006}Marois, C., et al., 2006, \apj, 641, 556
\bibitem[Marois et al.(2008)]{Marois2008}Marois, C., et al., 2008, Science, 322, 1348
\bibitem[Marois et al.(2010)]{Marois2011}Marois, C., et al., 2010, Nature, 468, 1080
\bibitem[Mohanty et al.(2007)]{Mohanty2007}Mohanty, S., et al., 2007, 657, 1064
\bibitem[Pecaut et al.(2012)]{Pecaut2012}Pecaut, M.~J., et al., 2012, \apj, 746, 154
\bibitem[Pecaut et al.(2013)]{Pecaut2013}Pecaut, M.~J., \& Mamajek, E.~E.\ 2013, \apjs, 208, 9
\bibitem[Ratzka et al.(2005)]{Ratzka2005}Ratzka, T., Kohler, R., Leinert, C., 2005, A\&A, 437, 611
\bibitem[Rameau et al.(2013)]{Rameau2013}Rameau, J., et al., 2013, \apj, 779, L26
\bibitem[Saumon et al.(2012)]{Saumon2012}Saumon, D., et al., 2012, \apj, 750, 74
\bibitem[Simon et al.(1995)]{Simon1995}Simon, M., et al., 1995, \apj, 443, 625
\bibitem[Slesnick et al.(2004)]{Slesnick2004}Slesnick, C., et al., 2004, \aj, 610, 1045
\bibitem[Stephens et al.(2009)]{Stephens2009}Stephens, D., et al., 2009, \apj, 702, 154
\bibitem[Yelda et al.(2010)]{Yelda2010}Yelda, S., et al., 2010, \apj, 725, 331
\end{thebibliography}
\end{document}